\documentclass[10pt,journal,compsoc]{IEEEtran}
\usepackage[pdftex]{graphicx}
\usepackage{epstopdf}
\usepackage[cmex10]{amsmath}
\usepackage{cases,amssymb}
\usepackage{subeqnarray,subfigure}
\usepackage{algorithmic}
\usepackage[linesnumbered,ruled,vlined]{algorithm2e}
\usepackage{cite,setspace,bm,enumerate}
\usepackage{amsmath,color}
\usepackage{amsfonts,caption}
\usepackage{url}
\usepackage{multirow,array}
\usepackage{bbding} 

\newtheorem{theorem}{\textbf{Theorem}}

\allowdisplaybreaks[3]

\begin{document}

\title{Dynamic Human Digital Twin Deployment at the Edge for Task Execution: A Two-Timescale Accuracy-Aware Online Optimization}%

\author{
Yuye Yang,
You Shi,
Changyan Yi,~\IEEEmembership{Member,~IEEE},
Jun Cai,~\IEEEmembership{Senior Member,~IEEE},
Jiawen Kang,\\
Dusit Niyato,~\IEEEmembership{Fellow,~IEEE},
and Xuemin (Sherman) Shen,~\IEEEmembership{Fellow,~IEEE}
 %
\IEEEcompsocitemizethanks{\IEEEcompsocthanksitem Y. Yang, Y. Shi and C. Yi are with the College of Computer Science and Technology, Nanjing University of Aeronautics and Astronautics,
 Nanjing, China. (E-mail: \{mryyy, shyou, changyan.yi\}@nuaa.edu.cn).
\IEEEcompsocthanksitem J. Cai is with the Department of Electrical and Computer Engineering, Concordia University, Montr\'{e}al, QC, H3G 1M8, Canada. (E-mail: jun.cai@concordia.ca).
\IEEEcompsocthanksitem J. Kang is with the School of Automation, Guangdong University of Technology, Guangzhou 510062, China (E-mail: kavinkang@gdut.edu.cn).
\IEEEcompsocthanksitem D. Niyato is with the School of Computer Science and Engineering, Nanyang Technological University, Singapore. (E-mail: dniyato@ntu.edu.sg).
\IEEEcompsocthanksitem X. (Sherman) Shen is with the Department of Electrical and Computer Engineering, University of Waterloo, Waterloo, ON N2L 3G1, Canada. (E-mail: sshen@uwaterloo.ca).
}%


}

\IEEEtitleabstractindextext{%

\begin{abstract}
Human digital twin (HDT) is an emerging paradigm that bridges physical twins (PTs) with powerful virtual twins (VTs) for assisting complex task executions in human-centric services. In this paper, we study a two-timescale online optimization for building HDT under an end-edge-cloud collaborative framework. As a unique feature of HDT, we consider that PTs' corresponding VTs are deployed on edge servers, consisting of not only generic models placed by downloading experiential knowledge from the cloud but also customized models updated by collecting personalized data from end devices. To maximize task execution accuracy with stringent energy and delay constraints, and by taking into account HDT's inherent mobility and status variation uncertainties, we jointly and dynamically optimize VTs' construction and PTs' task offloading, along with communication and computation resource allocations. Observing that decision variables are asynchronous with different triggers, we propose a novel two-timescale accuracy-aware online optimization approach (TACO). Specifically, TACO utilizes an improved Lyapunov method to decompose the problem into multiple instant ones, and then leverages piecewise McCormick envelopes and block coordinate descent based algorithms, addressing two timescales alternately. Theoretical analyses and simulations show that the proposed approach can reach asymptotic optimum within a polynomial-time complexity, and demonstrate its superiority over counterparts.
\end{abstract}

}

\maketitle
\IEEEdisplaynontitleabstractindextext
\IEEEpeerreviewmaketitle

\section{Introduction}\label{Intro}

\IEEEPARstart{H}{uman} digital twin (HDT) is defined as a paradigm that can vividly characterize the replication of each individual human in the virtual space while real-time reflecting its actual physical and mental status in the physical space \cite{okegbile2022human}, \cite{lin2022human}.
With the personalized status information maintained in a high-fidelity virtual environment, HDT can be regarded as a ``sandbox'', where complex tasks for human-centric services (e.g., activity recognition \cite{thamotharan2023human} and vital signal measurement \cite{bjornsson2020digital}) are able to be repeatedly simulated and tested, guiding the practical implementation.
Because of the large potential in assisting complex task execution with human-centric concerns, HDT has been envisioned as a key enabler for Metaverse, Healthcare 5.0, Society 5.0, etc., attracting significant attentions recently \cite{chen2023networking}.


Essentially, the HDT system consists of a number of physical twin (PT) and virtual twin (VT) pairs, where PT stands for the physical entity (i.e., human) and VT represents the corresponding virtual model \cite{okegbile2022edge}. Obviously, the successful realization of HDT largely depends on the well construction and management of VT, so as to provide fast-responsive interactions and high-accurate task execution for its paired PT. These requirements prompt the adoption of end-edge-cloud collaborative framework \cite{ren2019survey}, by which HDT can be built and operated at the network edge (while supported by both end devices and the cloud center), guaranteeing pervasive connectivities, customized services and low-delay feedbacks.
Although some preliminary efforts have been dedicated on studying similar problems, such as industrial digital twin construction at the edge \cite{jia2022accurate}, \cite{dong2019deep}, \cite{10301793} and service application deployment across edges \cite{wang2022online}, \cite{ouyang2018follow}, \cite{harutyunyan2020latency}, establishing HDT at the edge for assisting task execution particularly involves some fundamentally different and unique issues that remain unexplored but are of great importance.
On one hand, different from position-fixed industrial plants, PTs in the HDT system are highly mobile with unpredictable mobility patterns (including positional and postural variations), leading to potential instability of PT-VT connectivities \cite{okegbile2022human}. Therefore, to guarantee seamless PT-VT interactions, it is necessary to dynamically place the associated VT of each PT on the edge server (ES) that this PT may switch its access to (caused by the random mobility).
On the other hand, unlike generalized applications requesting encapsulated services, PTs in the HDT system are extremely personalized and their status may vary frequently by uncertain external factors or physiological state changes, resulting in the potential inconsistency between each PT and its associated VT \cite{wang2024human}. Hence, to keep up-to-date high-fidelity VTs, it is necessary to keep the associated VT on the ES updated in a real-time manner (especially for the customized part).

Nevertheless, meeting all aforementioned requirements are very challenging because of the following reasons.
\begin{itemize}
\item[1)] To enhance the accuracy of complex task execution assisted by HDT, it is required to construct fine-grained VTs on ESs (by both dynamically place generic models via the cloud and update customized toppings via sensors worn on PTs). However, the data brought by VT constructions can be massive \cite{lauer2022designing}, inherently increasing the service delay and energy consumption, and thus the data size of generic model placement and customized model update should be carefully optimized for striking a balance between accuracy and cost. On top of this, considering that the ultimate goal is to assist the task execution, how to timely and efficiently offloading tasks from PTs to ESs (containing associated VTs) should also be jointly considered with VT constructions, because both of them share the same communication and computation resources.
\item[2)] Since HDT is time-varying evolutionalized with uncertain PT-VT mobility and status variations, the system optimization has to be conducted online while the statistics of future information (related to human activities) may be hard to obtain, if not impossible \cite{shi2023service}. Moreover, the dynamic placement of generic VT models is triggered by PTs' mobility and access handover among different ESs, which usually happens over a long time period. In contrast, the dynamic update of customized VT models and the complex task offloading are triggered by PTs' status variations, which may need to be adapted in a much higher frequency. These indicate that such actions for system optimization should be performed asynchronously in different timescales.
\end{itemize}

To tackle the aforementioned difficulties, in this paper, we propose a novel two-timescale accuracy-aware online optimization approach for building HDT in assisting complex task execution under an end-edge-cloud collaborative framework.
Specifically, we consider that each PT's associated VT (deployed on the ES) consists of a generic model placed by downloading experiential knowledge (e.g., feature parameters and weights) from the cloud center and a customized model updated by uploading personalized data (e.g., behavior characteristics) from sensors worn on the PT.
With the objective of maximizing the average accuracy of complex task execution assisted by HDT under stringent energy and delay constraints, and by taking into account the system uncertainties (e.g., random mobility and status variations), we formulate a two-timescale online optimization problem.
Particularly, we aim to dynamically optimize i) large-timescale decisions, including the granularity of each PT's experiential knowledge for placing its generic VT model and the ES access selection of each PT, and ii) small-timescale decisions, including the amount of each PT's personalized data for updating its customized VT model, task offloading decision (i.e., local computing or edge computing) for each task, and communication and computation resource allocations of each ES.
To this end, we develop a novel two-timescale accuracy-aware online optimization approach (TACO) based on the improved Lyapunov optimization. Specifically, the long-term problem is first decomposed into a series of short-term deterministic subproblems with different timescales, and then an alternating algorithm is proposed, integrating piecewise McCormick envelopes (PME) and block coordinate descent (BCD) based methods, for iteratively solving these subproblems. Theoretical analyses show that the proposed approach can produce an asymptotically optimal outcome with a polynomial-time complexity.

The main contributions of this paper are summarized in the following.
\begin{itemize}
  \item To the best of our knowledge, we are the first to study the HDT deployment at the network edge for assisting human-centric task execution by formulating a two-timescale accuracy-aware online optimization problem, which jointly optimizes VTs' construction (including dynamic generic model placement and customized model update) and PTs' task offloading together with the management of PT-ES access selection and corresponding communication and computation resource allocations.
  \item We propose a novel approach, called TACO, which first decomposes the long-term optimization problem into multiple instant ones. Then, we leverage PME and BCD based algorithms for alternately solving the decoupled subproblems in the large-timescale and small-timescale, respectively.
  \item We theoretically analyze performance of the proposed TACO approach by rigorously deriving the gap to optimum and the computational complexity in the closed-form. Furthermore, extensive simulations show that the proposed TACO approach can outperform counterparts in terms of improving the HDT-assisted task execution accuracy, and reducing the service response delay and overall system energy consumption.
\end{itemize}

The rest of this paper is organized as follows. Section \ref{RW} reviews the recent related work and highlights the novelties of this paper. Section \ref{SM} describes the considered system model and the problem formulation.
In Section \ref{Algorithm}, the two-timescale accuracy-aware online optimization approach, i.e., TACO, is proposed and analyzed theoretically.
Simulation results are presented in Section \ref{PE}, followed by the conclusion in Section \ref{Conclusion}.


\section{Related Work}\label{RW}

As one of the key enabler for emerging applications, HDT has recently drawn a lot of research attentions from both academia and industry. For example, Lee et al. in \cite{lee2023meta} proposed a large-scale HDT construction framework on the cloud server integrated with a synchronization mechanism to reduce system overall data transmission cost. Zhong et al. in \cite{zhong2023construction} introduced a bidirectional long short-term memory based algorithm in designing high-fidelity HDT model with multimodel data on the cloud platform. In \cite{liu2019novel}, Liu et al. developed a cloud HDT based healthcare system by optimizing a patient-data processing workflow to improve the quality of personal health management. However, these papers were mainly restricted to constructing HDT sorely on the cloud server, ignoring the potential of utilizing network edge resources for empowering HDT with the capability of providing pervasive, customized and low-delay services.

While deploying HDT at the network edge has rarely been investigated, some researchers have dedicated in studying the general DT construction at the edge. Dong et al. in \cite{dong2019deep} proposed a deep learning algorithm for constructing DTs of the mobile edge network, aiming to minimize the normalized energy consumption through the optimization of user associations, resource allocations and offloading probabilities. Zhang et al. in \cite{10301793} formulated a DT adaptive placement and transfer problem to minimize the DT synchronization delay, which were then solved by the double auction based optimization. Nevertheless, these papers considered that DTs were constructed on fixed locations or placed following pre-known mobility patterns, making them unsuitable for HDT with human-centric features, where PTs are highly mobile with unpredictability.
Another stream of related works have been conducted on general mobile service application deployment across edges. For instance, Wang et al. in \cite{wang2022online} developed a user-centric learning-driven method for deploying and migrating delay-aware service applications to minimize the total service delay of mobile users. Ouyang et al. \cite{ouyang2018follow} formulated a dynamic service deployment problem with the objective of minimizing the user-perceived delay under the uncertain user mobility. However, in these papers, service applications were assumed to have limited and encapsulated types, meaning that they are not customized and do not need to be updated, which largely differ from those of HDT (where on-demand evolution is essential).

To guarantee the long-term performance in online problems, Lyapunov optimization method has been widely recognized as an efficient approach \cite{lin2021stochastic}, \cite{ding2021potential}, \cite{shi2022joint}, yet most of existing solutions were restricted to problems with decisions in the single timescale only. Recently, some preliminary studies \cite{he2022two}, \cite{ma2020leveraging} have delved into designing two-timescale Lyapunov methods, by which the original problem was decomposed and further decoupled into subproblems in two different timescales independently and then optimized separately. Besides these, in \cite{shi2023service}, \cite{zhou2022two}, alternating algorithms were developed in addition to the Lyapunov framework to tackle subproblems in two timescales with coupled relationships but are both convex (or can be easily converted into convex forms). However, these solutions cannot be directly applied in this paper because the considered subproblems (after the decomposition) are not only tightly coupled but also highly non-convex.

In summary, different from all the existing works, this paper proposes a novel two-timescale accuracy-aware online optimization approach to jointly optimize the HDT deployment (i.e., generic placement and customized update of VT model) and task offloading under an end-edge-cloud collaborative framework, where the novelty lies in not only the system model but also the solution.

\section{System Model and Problem Formulation}\label{SM}
In this section, we first provide a system overview on how HDT is deployed and functioned on ESs. Then, to be more specific, the generic VT model placement in the large-timescale and the customized VT model update together with the task offloading in the small-timescale are described. After that, aiming to enhance the accuracy of complex task execution assisted by HDT, a two-timescale online optimization problem is formulated.

\subsection{System Overview}
\begin{figure}[!t]
\begin{center}
\includegraphics[width=8.5 cm]{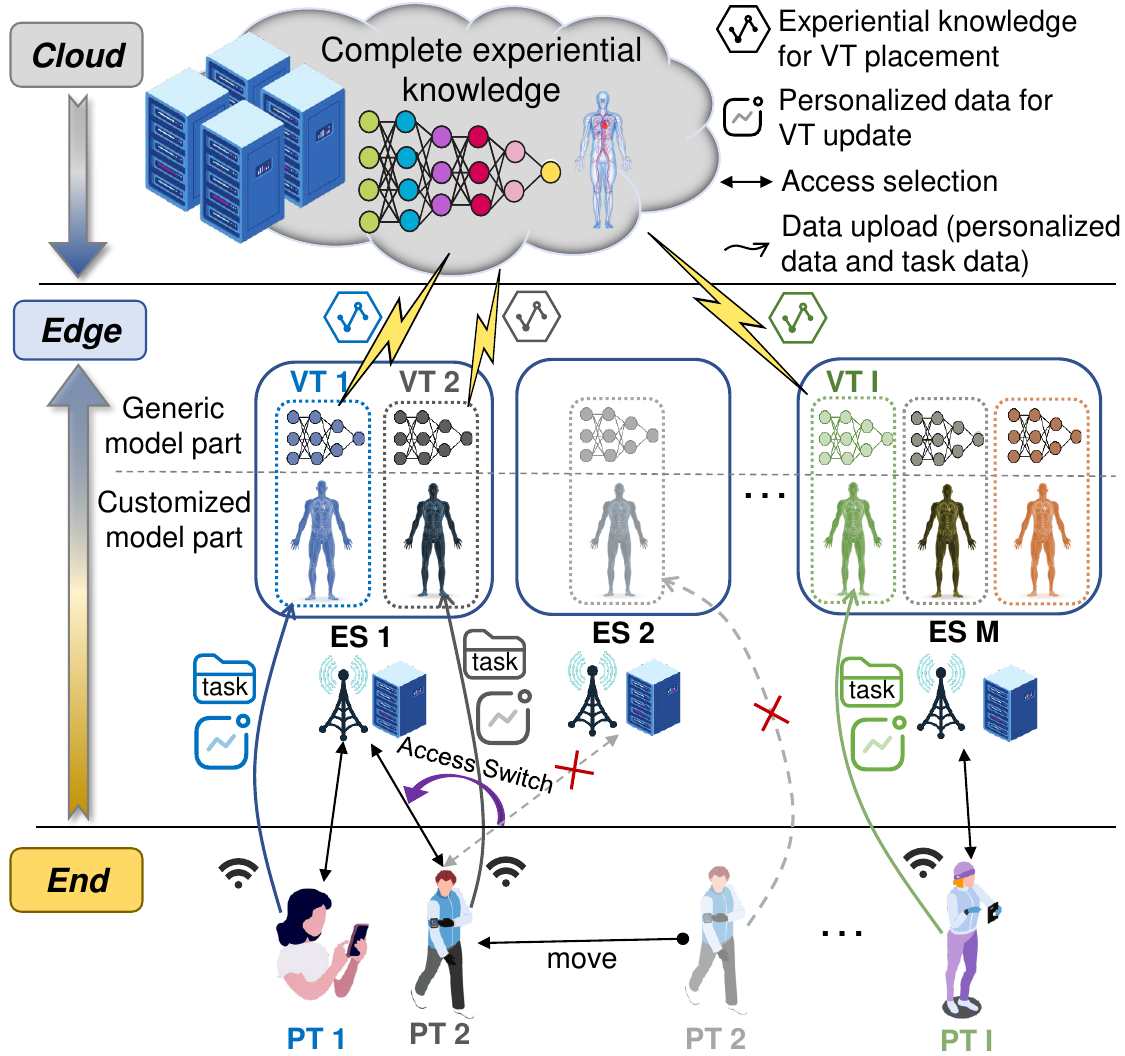}
  \caption{The end-edge-cloud collaborative HDT system.}
  \label{model2}
\end{center}
\end{figure}
Consider an HDT system building upon an end-edge-cloud collaborative framework, as illustrated in Fig. \ref{model2}, consisting of a set of end users (regarded as PTs) $\mathcal{I}$ with cardinality of $\mid \mathcal{I}\mid=I$, multiple geographically distributed ESs denoted as $\mathcal{M}$ with $\mid \mathcal{M}\mid=M$, and a cloud center (acting as the central controller).
PTs (roaming around) generate streams of complex tasks which require the construction of exclusive VT models (forming one-to-one PT-VT pairs) at the edge to assist their task executions. Each VT should be deployed on the ES to which its associated PT may access for offering high-quality and low-delay services. Note that one ES can host multiple VT models for different PTs, and the corresponding communication and computation resources are shared among them. Furthermore, the construction of a high-fidelity VT on the ES consists of two main procedures, i.e., generic model placement and customized model update \cite{shengli2021human}. For each VT, the generic part of the model is obtained by downloading the experiential knowledge with a selected granularity\footnote{Selecting a large (small) granularity of experiential knowledge for generic model placement may increase (decrease) the fidelity of the VT model, while also introducing a large (small) amount of data to be transferred, resulting in high (low) service delay and energy consumption.} from the cloud center, and the target ES for its placement is determined following the access selection of the associated PT. By contrast, the customized part of each VT model is updated by uploading the personalized data with an optimized data size\footnote{The required size of personalized data for customized model update affects not only the fidelity of the VT model, but also the uplink communication and computation resource allocations among different PT-VT pairs and between their data transmission and task offloading.} obtained from sensors worn on the associated PT. After VT establishment, PTs' tasks can be either transmitted (offloaded) to VTs deployed on the ES or processed locally, depending on the demands of task execution accuracy versus the requirements of service delay and energy consumptions.
It is worth noting that although VT models are not able to be built locally, PTs' tasks can be executed by running offline service applications pre-installed on PTs, which are much less powerful but do not require to be real-time updated.


\begin{figure}[!t]
\begin{center}
\includegraphics[width=8.5 cm]{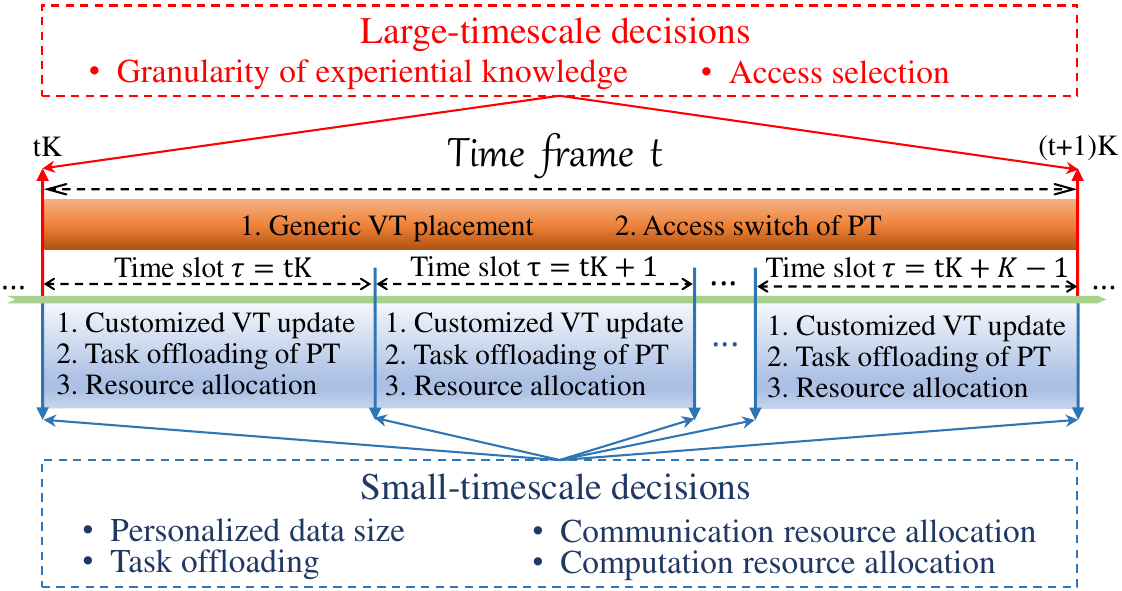}
  \caption{Two-timescale online optimization framework.}
  \label{model3}
\end{center}
\end{figure}

In practice, the dynamic placement of generic VT model and ES access handover of each PT commonly happen in a low frequency (i.e., over a large-timescale)\footnote{Particularly, frequently downloading experiential knowledge and switching access points can lead to large configuration costs, including the dramatic increase of delays and energy consumptions \cite{chen2023networking}.}, while the update of customized VT models and task offloading together with corresponding communication and computation resource allocations require immediate and frequent adaptation to accommodate the status variation of PT and its task generations \cite{lin2022human}.
To this end, we define that in the considered online optimization framework, the access selection of each PT and the granularity of experiential knowledge for its generic VT model placement are decided in the large-timescale, while the amount of personalized data for its customized VT model update, task offloading, communication and computation resource allocations are decided in the small-timescale, as shown in Fig. \ref{model3}.
Specifically, the timeline is segmented into $T \in \mathbb{N}^{+}$ coarse-grained time frames, and each frame can be further divided into a combination of $K \in \mathbb{N}^{+}$ fine-grained time slots.
Let $t\in\mathcal{T}=\{0,1, \ldots,T-1\}$ be the index of the $t$-th time frame, and define $\tau \in \mathcal{T}_{t}=\{tK,tK+1, \ldots, tK+K-1\}$ as the index of the $\tau$-th time slot in the $t$-th time frame.

Overall speaking, we target to optimize the long-term system performance (i.e., the average accuracy of complex task execution assisted by HDT under stringent delay and energy consumption constraints) by determining i) which ES should be selected to access for each PT and what level of granularity of the experiential knowledge should be chosen for placing its generic VT model on the ES in each time frame, and ii) how large the personalized data is required for updating the customized part of each VT model and whether the task of each PT should be processed locally or on the ES (deployed with its associated VT) together with communication and computation resource allocations in each time slot, in an online manner.
For convenience, all important notations in this paper are listed in Table \ref{tab1}.


\begin{table}[!t]
\footnotesize
\renewcommand{\arraystretch}{1.1}
\caption{IMPORTANT NOTATIONS IN THIS PAPER}
\label{tab1}
\centering
\begin{tabular}{p{29pt}|p{195pt}}
  \hline
  \bfseries Symbol & \bfseries Meaning \\
  \hline\hline
  $A_i(\tau)$ & task execution accuracy for PT $i$ in time slot $\tau$\\
  $a_{i,m}(t)$ & access selection of PT $i$ in time frame $t$\\
  $B_m$ & total bandwidth resource for accessing ES $m$\\
  $b_{i}(\tau)$ & bandwidth resource allocation of PT $i$ in time slot $\tau$\\
  $D_i(t)$ & total experiential knowledge of PT $i$ in time frame $t$\\
  $E_{i}^{dl}(t)$ & energy consumption of downloading the experiential knowledge of PT $i$ in time frame $t$\\
  $E_{i,m}^{pl}(t)$ & energy consumption of placing generic VT model $i$ in ES $m$ in time frame $t$\\
  $E_{i,m}^{ul}(\tau)$ & energy consumption of uploading the personalized data from PT $i$ to ES $m$ in time slot $\tau$\\
  $E_{i,m}^{ud}(\tau)$ & energy consumption of updating the customized VT model $i$ on the ES $m$ in time slot $\tau$\\
  $E_{i,m}^{ofld}(\tau)$ & energy consumption of transmitting task data of PT $i$ to ES $m$ in time slot $\tau$\\
  $E_{i}^{exec}(\tau)$ & energy consumption of executing PT $i$'s task\\
  $F_i$ & total computation resource of PT $i$'s local device\\
  $F_m$ & total computation resource of ES $m$ for all VTs\\
  $f_{i}^{}(\tau)$ & computation resource allocation of PT $i$ in time slot $\tau$\\
  $\mathcal{I}$ & set of end users (PTs) / corresponding VTs \\
  $\mathcal{M}$ & set of ESs  \\
  $r_{i,m}(\tau)$ & transmission rate between PT $i$ and ES $m$ in time slot $\tau$\\
  $S_i(\tau)$ & total personalized data generated by PT $i$ in time slot $\tau$\\
  $T_{i}^{dl}(t)$ & delay of downloading the experiential knowledge of PT $i$ in time frame $t$\\
  $T_{i,m}^{pl}(t)$ & delay of placing generic VT model $i$ on ES $m$\\
  $T_{i,m}^{ul}(\tau)$ & delay of uploading the personalized data from PT $i$ to ES $m$ in time slot $\tau$\\
  $T_{i,m}^{ud}(\tau)$ & delay of updating customized VT model $i$ on the ES $m$\\
  $T_{i,m}^{ofld}(\tau)$ & delay of transmitting task data of PT $i$ to ES $m$\\
  $T_{i}^{exec}(\tau)$ & delay of task execution of PT $i$ in time slot $\tau$\\
  $V$ & Lyapunov control parameter\\
  $x_{i}(t)$ & experiential knowledge granularity for generic model placement of PT $i$ in time frame $t$\\
  $y_{i}^{}(\tau)$ & personalized data size for customized model update of PT $i$ in time slot $\tau$\\
  $z_i(\tau)$ & task offloading of PT $i$ in time slot $\tau$\\
    \hline
\end{tabular}
\end{table}

\subsection{Generic VT Model Placement}
Since PTs are mobile, to construct VT for each of them at the network edge so as to enable seamless PT-VT interactions, the generic VT model should be re-displaced on the ES that its associated PT switches its access to in each time frame $t\in\mathcal{T}$.
Denote $a_{i,m}(t)\in\{0,1\}$ as the access selection decision in the large-timescale indicating whether PT $i\in\mathcal{I}$ selects to access ES $m\in\mathcal{M}$ or not in time frame $t \in \mathcal{T}$, i.e., $a_{i,m}(t)=1$ if PT $i\in\mathcal{I}$ connects to ES $m\in\mathcal{M}$, and $a_{i,m}(t)=0$ otherwise.
Obviously, we should have $\sum_{m\in\mathcal{M}} a_{i,m}(t)\leq 1$, meaning that PT $i$ cannot access multiple ESs simultaneously \cite{shi2023service}.
For each PT $i\in\mathcal{I}$, its generic VT model is placed on the accessed ES by downloading a certain granularity of experiential knowledge from the cloud center.
We define that the full experiential knowledge of each PT $i\in\mathcal{I}$ for its generic VT model placement has a total size of $D_i(t)$, and denote $x_{i}(t) \in [0,1]$ as its decision of granularity in time frame $t \in \mathcal{T}$. Then, the data size of downloading the experiential knowledge for placing PT $i$'s VT model in time frame $t\in\mathcal{T}$ can be represented as $x_{i}(t)D_i(t)$.
Based on these, the corresponding delay and energy consumption of downloading such experiential knowledge can be respectively expressed as\footnote{Note that even for a special case that PT $i$'s access selection remains unchanged, it is still necessary to periodically replace its generic VT model on the same ES, because the experiential knowledge may experience ``data drift'' over the time \cite{xie2024dual}.}
\begin{equation}
T_{i}^{dl}(t)=\sum_{m\in\mathcal{M}}a_{i,m}(t)\frac{x_{i}(t)D_i(t)}{r^c}, \forall i\in\mathcal{I}, \forall t\in\mathcal{T},
\end{equation}
\begin{equation}
E_{i}^{dl}(t)=T_{i}^{dl}(t)p^c, \forall i\in\mathcal{I}, \forall t\in\mathcal{T},
\end{equation}
where $r^c$ and $p^c$ stand for the downlink transmission rate and unit transmission power from the cloud center to each ES, respectively, which are both considered as constants \cite{zhou2022two}, \cite{mohammadi2018uplink}.

To exploit this experiential knowledge, each ES has to allocate a proportion of its computation resource for completing the generic VT model placement at the beginning of each time frame.
The delay of doing this for PT $i\in\mathcal{I}$ on ES $m\in\mathcal{M}$ in time frame $t \in \mathcal{T}$ can be calculated as
\begin{equation}
T_{i,m}^{pl}(t)=\frac{a_{i,m}(t)x_{i}(t)D_i(t) C_m}{f_{i}(tK)F_m},
\end{equation}
where $C_m$ is the number of CPU cycles required for ES $m$ to process a unit of data, $F_m$ is the CPU speed (measured by cycles/s) of ES $m$, and $f_{i}(tK)\in(0,1]$ represents the ratio of computation resource allocated to PT $i$ for its VT construction at the beginning of time frame $t$ (i.e., the first time slot of the frame with $\tau=tK$).
Referring to the energy model widely used in CMOS circuits\cite{yi2020queueing}, the energy consumption of constructing the generic VT model of the associated PT $i\in\mathcal{I}$ on the ES $m\in\mathcal{M}$ in each time frame $t \in \mathcal{T}$ can be calculated as
\begin{equation}
E_{i,m}^{pl}(t)=\rho_m f_{i}(tK)(F_m)^3 T_{i,m}^{pl}(t),
\end{equation}
where $\rho_m$ is the effective switched capacitance of ES $m$ depending on its chip architecture.

\subsection{Customized VT Model Update}

Since PTs are personalized and their status may vary frequently due to uncertain external or internal factors, to guarantee the timeliness and high-fidelity of VTs on ESs, the customized VT model of each PT should be updated in each time slot $\tau \in \mathcal{T}_{t}$.
Let $S_i(\tau)$ be the total amount of personalized data generated by PT $i \in\mathcal{I}$, and define $y_i(\tau)\in[0,1]$ as the percentage of personalized data chosen to be uploaded in time slot $\tau \in \mathcal{T}_{t}$. Then, the size of personalized data uploaded for updating PT $i$'s customized VT model in time slot $\tau\in\mathcal{T}_t$ can be expressed as $y_i(\tau)S_i(\tau)$.

Within each time slot $\tau\in \mathcal{T}_t$, we denote the location of PT $i\in\mathcal{I}$ as $(x_i(\tau), y_i(\tau))$, which is a state information following its random mobility pattern, and let $(x_m,y_m)$ be the fixed location of each ES $m\in\mathcal{M}$.
The distance between any PT $i\in\mathcal{I}$ and ES $m\in\mathcal{M}$ can then be calculated as $S_{i,m}(\tau)=\sqrt{\left(x_i(\tau)-x_m\right)^2+\left(y_i(\tau)-y_m\right)^2}$, and according to Shannon-Hartley formula, the transmission rate from PT $i\in\mathcal{I}$ to its accessed ES $m\in\mathcal{M}$ is written as
\begin{equation}
\hspace{-5.2pt}r_{i,m}(\tau)\hspace{-2.5pt}=\hspace{-2.3pt} a_{i,m}(t)\hspace{-1pt}b_{i}(\tau)\hspace{-1pt}B_{m}\hspace{-2pt} \log(1\hspace{-2pt}+\hspace{-1.5pt}\frac{(S_{i,m}(\tau))^\theta p_i|h_{i ,m}(\tau)|^2}{ N_0 b_{i}(\tau)B_m}),
\end{equation}
where $b_i(\tau)\in(0,1]$ is the proportion of communication resource allocated to PT $i\in\mathcal{I}$ in time slot $\tau\in\mathcal{T}_t$,
$h_{i,m}(\tau)$ is the fading amplitude between PT $i\in\mathcal{I}$ and ES $m\in\mathcal{M}$ in time slot $\tau \in \mathcal{T}_{t}$ (modeled as a circularly symmetric complex Gaussian random variable \cite{goldsmith2005wireless}),
$B_m$ is the communication bandwidth of ES $m\in\mathcal{M}$,
$N_0$ is the spectral density of the channel noise power,
$p_{i}$  is the pre-determined transmission power of PT $i\in\mathcal{I}$,
and $\theta \geq 2$ is the path loss exponent \cite{goldsmith2005wireless}.
Correspondingly, the delay and energy consumption of PT $i\in\mathcal{I}$ in uploading the personalized data with size $y_i(\tau)S_i(\tau)$ for updating its VT on ES $m\in\mathcal{M}$ in each time slot $\tau\in\mathcal{T}_t$ can be respectively expressed as
\begin{equation}
T_{i,m}^{ul}(\tau)=\frac{y_{i}(\tau)S_i(\tau)}{r_{i,m}(\tau)}, \forall i\in\mathcal{I}, \forall m\in\mathcal{M}, \forall \tau\in\mathcal{T}_t,
\end{equation}
\begin{equation}
E_{i,m}^{ul}(\tau)=T_{i,m}^{ul}p_{i}, \forall i\in\mathcal{I}, \forall m\in\mathcal{M}, \forall \tau\in\mathcal{T}_t.
\end{equation}

To utilize this personalized data, each ES has to allocate a proportion of its computation resource for completing the customized VT model update in each time slot.
The delay of doing this for PT $i\in\mathcal{I}$ on ES $m\in\mathcal{M}$ in time slot $\tau \in \mathcal{T}_{t}$ can be calculated as
\begin{equation}
T_{i,m}^{ud}(\tau)=\frac{a_{i,m}(t)y_{i}(\tau)S_i(\tau) C_m}{f_{i}(\tau)F_m},
\end{equation}
where $f_i(\tau)\in(0,1]$ indicates the proportion of computation resource allocated to PT $i$ for its VT update in each time slot $\tau\in[tK,tK+K-1]$.
Besides, the corresponding energy consumption can be calculated as
\begin{equation}
E_{i,m}^{ud}(\tau)=\rho_mf_{i}(\tau)(F_m)^3T_{i}^{ud}(\tau).
\end{equation}

\subsection{HDT-Assisted Task Execution}
Let $\lambda_i(\tau)$ be the data size of the complex task produced by PT $i\in\mathcal{I}$ in each time slot $\tau \in \mathcal{T}_t$, which is allowed to follow a general random distribution.
Denote the task offloading decision of PT $i\in\mathcal{I}$ in time slot $\tau \in \mathcal{T}_t$ as $z_i(\tau)\in\{0,1\}$, i.e., $z_i(\tau)=1$ if PT $i$ offloads the task to its VT on the ES for assistance, and $z_i(\tau)=0$ if PT $i$ processes it locally.
The delay and energy consumption of offloading/transmitting such task from PT $i\in\mathcal{I}$ to its associated VT deployed on the ES $m\in\mathcal{M}$ in time slot $\tau \in \mathcal{T}_{t}$ can be respectively expressed as
\begin{equation}
T_{i,m}^{ofld}(\tau)=\frac{\lambda_i(\tau)}{r_{i,m}(\tau)},
\end{equation}
\begin{equation}
E_{i,m}^{ofld}(\tau)=T_{i,m}^{ofld}p_{i}.
\end{equation}

Considering the possibility of both edge and local processing, the delay of HDT-assisted task execution of PT $i\in\mathcal{I}$ in time slot $\tau \in \mathcal{T}_{t}$ can be calculated as
\begin{equation}
\begin{aligned}
& T_{i}^{exec}(\tau)=\sum_{m\in\mathcal{M}}a_{i,m}(t) z_i(\tau)\frac{\lambda_i(\tau) C_m}{f_{i}(\tau)F_m}\\
& +(1-z_i(\tau))\frac{\lambda_i(\tau)C_i}{F_i},
\end{aligned}
\end{equation}
where $C_i$ is the number of CPU cycles required for PT $i$ to locally process a unit of data, and $F_i$ denotes its CPU speed (measured by cycles/s).
Besides, the corresponding energy consumption can be calculated as
\begin{equation}
\begin{aligned}
& E_{i}^{exec}(\tau)=\sum_{m\in\mathcal{M}}a_{i,m}(t)z_i(\tau)\rho_m(F_m)^2\lambda_i(\tau) C_m\\
& +(1-z_i(\tau))\rho_i(F_i)^2\lambda_i(\tau)C_i.
\end{aligned}
\end{equation}

With the help of HDT at the edge, the accuracy of executing each task from PT $i\in\mathcal{I}$ in each time slot $\tau\in\mathcal{T}_t$ can be defined as
\begin{equation}\label{AccuracyEqu}
A_i(\tau)=z_i(\tau)g_i^{edge}(d_i(\tau))+(1-z_i(\tau))g_i^{local},
\end{equation}
where $g_i^{local}$ represents the task execution accuracy of local processing on PT $i$ itself, and $g_i^{edge}(d_i(\tau))$ stands for the task execution accuracy of edge computing for PT $i$ depending on the total size of data used for its VT construction, i.e., $d_i(\tau)=x_{i}(t)D_i(t) + y_{i}(\tau)S_i(\tau),\forall\tau\in\mathcal{T}_t,\forall t\in\mathcal{T}$, consisting of both experiential knowledge for generic VT model placement in the large-timescale and personalized data for customized VT model update in the small-timescale. Note that $g_i^{edge}(d_i(\tau))$ is a mapping function that can be obtained via empirical studies or experiments \cite{wu2020accuracy}.

Similar to \cite{yi2020queueing}, \cite{wu2017dynamic}, we ignore overheads induced by computing outcomes feedback and system control signals. This is because the size of computing outcomes and control signals are much smaller than that of the input data. Technically, these overheads can be seen as small constants \cite{yi2021workload}, which will not affect our analyses.

\subsection{Problem Formulation}\label{PF}
In summary, the total response delay for all tasks of PT $i\in\mathcal{I}$ in each time frame $t \in \mathcal{T}$ can be derived as
\begin{equation}\label{toldelay}
\begin{aligned}
& T_i^{tol}(t)=T_{i}^{dl}(t)+\sum_{m\in\mathcal{M}}T_{i,m}^{pl}(t)+\sum_{\tau\in\mathcal{T}_t}[z_i(\tau)\sum_{m\in\mathcal{M}}\\
& (T_{i,m}^{ul}(\tau)+T_{i,m}^{ud}(\tau)+T_{i,m}^{ofld}(\tau))+T_{i}^{exec}(\tau)],
\end{aligned}
\end{equation}
and the system overall energy consumption in each time frame $t \in \mathcal{T}$ can be derived as
\begin{align}\label{tolenergy}
& E^{tol}(t)=\sum_{i\in\mathcal{I}}(E_{i}^{dl}(t)+\sum_{m\in\mathcal{M}}E_{i,m}^{pl}(t))+\sum_{i\in\mathcal{I}}\sum_{\tau\in\mathcal{T}_t}[z_i(\tau) \nonumber\\
& \sum_{m\in\mathcal{M}}(E_{i,m}^{ul}(\tau)+E_{i,m}^{ud}(\tau)+E_{i,m}^{ofld}(\tau))+E_{i}^{exec}(\tau)].
\end{align}
To evaluate the core value of building HDT at the network edge, we take the long-term average accuracy of complex task execution assisted by the considered end-edge-cloud collaborative HDT system over all time frames as the performance measurement, which can be expressed as
\begin{equation}
\mathcal{A}=\frac{1}{TK}\sum_{t\in\mathcal{T}}\sum_{\tau\in\mathcal{T}_t}\sum_{i\in\mathcal{I}}A_i(\tau).
\end{equation}

With the objective of maximizing $\mathcal{A}$ while ensuring that the response delay for all tasks of each PT $i\in\mathcal{I}$ and the overall system energy consumption do not exceed certain thresholds, we formulate a two-timescale online optimization problem by jointly optimizing $a_{i,m}(t)$ and $x_{i}(t)$, denoted in short by a set of large-timescale decision variables $\mathcal J_i^A(t)= \{a_{i,m}(t),x_{i}(t) \}$, for each PT $i$ in any time frame $t \in \mathcal{T}$, and $y_{i}^{}(\tau)$, $b_{i}(\tau)$, $f_{i}(\tau)$ and $z_{i}(\tau)$, denoted in short by a set of small-timescale decision variables $\mathcal J_i^B(\tau)=\{b_{i}(\tau),y_{i}^{}(\tau), f_{i}^{}(\tau),z_i(\tau)\}$, for each PT $i$ in any time slot $\tau \in \mathcal{T}_{t}$.

Mathematically, such a two-timescale online optimization problem can be formulated as
\begin{subequations}
\begin{align}
& \mathcal{P}_1: \max _{\mathcal J_i^A(t),\mathcal J_i^B(\tau)} \lim _{t\rightarrow \infty} \mathcal{A} \nonumber\\
& \text { s.t. } \sum_{m\in\mathcal{M}} a_{i,m}(t)\leq 1,\forall i\in\mathcal{I},\forall t\in\mathcal{T}, \label{C1}\\
& \sum_{i \in \mathcal{I}}a_{i,m}(t)b_{i}(\tau)\leq 1,\forall m\in\mathcal{M},\forall t\in\mathcal{T},\tau \in \mathcal{T}_t, \label{C3}\\
& \sum_{i\in\mathcal{I}}a_{i,m}(t)f_{i}(\tau)\leq 1,\forall m\in\mathcal{M},\forall t\in\mathcal{T},\tau\in\mathcal{T}_t, \label{C4}\\
& \lim_{T\rightarrow \infty}\frac{1}{T}\sum_{t\in\mathcal{T}} T_i^{tol}(t)\leq T_i^{max},\forall i\in\mathcal{I}, \label{LongTermDelay} \\
& \lim_{T\rightarrow \infty}\frac{1}{T}\sum_{t\in\mathcal{T}} E^{tol}(t)\leq E^{max},\label{LongTermEnergy}
\end{align}
\end{subequations}
where constraint (\ref{C1}) guarantees that one PT can connect to at most one ES in each time frame $t\in\mathcal{T}$;
constraints (\ref{C3}) and (\ref{C4}) restrict that the communication and computation resource allocation should be less than the total capacities of each ES in any time slot $\tau\in\mathcal{T}_t$;
constraints (\ref{LongTermDelay}) and (\ref{LongTermEnergy}) are the long-term average task response delay and system overall energy consumption constraints (in which $T_i^{max}$ and $E^{max}$ represent the pre-determined response delay threshold of each PT $i \in \mathcal{I}$ and the overall system energy consumption threshold, respectively).

Obviously, solving problem $\mathcal{P}_1$ directly is very challenging because
$i)$ PTs' mobilities and status variations, due to human-related characteristics, are highly unpredictable, meaning that it is extremely hard, if not impossible, to obtain system statistics in advance, which necessitates the design of an online optimization algorithm;
$ii)$ although Lyapunov optimization \cite{neely2022stochastic} is well-known as an effective method to solve such an online problem in general, decision variables in different timescales (i.e., $\mathcal J_i^A(t)$ and $\mathcal J_i^B(\tau)$) are tightly coupled in not only the objective function but also constraints (\ref{LongTermDelay}) and (\ref{LongTermEnergy}); and $iii)$ all constraints include discrete decision variables (i.e., $a_{i,m}(t)$ or $z_i(\tau)$), and constraints (\ref{LongTermDelay}) and (\ref{LongTermEnergy}) are both non-convex.
These indicate that $\mathcal{P}_1$ is a two-timescale online non-convex mixed integer programming problem, which must be NP-hard.

\section{A Two-Timescale Online Optimization Approach (TACO)}\label{Algorithm}
In this section, we propose a novel approach, namely TACO, for jointly optimizing VT construction and task offloading in the considered end-edge-cloud collaborative HDT system.
Specifically, we first reformulate the two-timescale problem by distributing the task response delay and system overall energy consumption of each time frame $t\in \mathcal{T}$ into each of its contained time slot $\tau\in \mathcal{T}_t$. Then, we decompose the problem into multiple short-term deterministic subproblems of different timescales with the help of Lyapunov optimization method but with a two-timescale extension. After that, we introduce an alternating algorithm integrating PME and BCD methods to solve subproblems in the large-timescale and small-timescale, respectively.

\subsection{Problem Reformulation}
Observed from problem $\mathcal{P}_1$ that the delay and energy consumptions caused by the generic VT model placement are on the large-timescale, while those caused by customized VT model update and task execution are on the small-timescale. To facilitate the solution, we evenly distribute the task response delay and system overall energy consumption in each time frame $t\in\mathcal{T}$ into all $\mid \mathcal{T}_t\mid=K$ time slots within this frame, which yields
\begin{equation}\label{toldelayslot}
\begin{aligned}
& T_i^{tol}(\tau)=(T_{i}^{dl}(t)+\sum_{m\in\mathcal{M}}T_{i,m}^{pl}(t))/K+\sum_{\tau\in\mathcal{T}_t}[z_i(\tau)\\
& \sum_{m\in\mathcal{M}}(T_{i,m}^{ul}(\tau)+T_{i,m}^{ud}(\tau)+T_{i,m}^{ofld}(\tau))+T_{i}^{exec}(\tau)],
\end{aligned}
\end{equation}
\begin{align}\label{tolenergyslot}
& E^{tol}(\tau)=\sum_{i\in\mathcal{I}}(E_{i}^{dl}(t)+\sum_{m\in\mathcal{M}}E_{i,m}^{pl}(t))/K+\sum_{i\in\mathcal{I}}\sum_{\tau\in\mathcal{T}_t}[z_i(\tau) \nonumber\\
& \sum_{m\in\mathcal{M}}(E_{i,m}^{ul}(\tau)+E_{i,m}^{ud}(\tau)+E_{i,m}^{ofld}(\tau))+E_{i}^{exec}(\tau)].
\end{align}
Substituting (\ref{toldelayslot}) and (\ref{tolenergyslot}) into (\ref{LongTermDelay}) and (\ref{LongTermEnergy}) of problem $\mathcal{P}_1$, we have
\begin{subequations}
\begin{align}
& \mathcal{P}_2: \max _{\mathcal J_i^A(t),\mathcal J_i^B(\tau)} \lim _{t\rightarrow \infty} \mathcal{A} \nonumber\\
& \text { s.t. } (\ref{C1}), (\ref{C3}), (\ref{C4}),\nonumber\\
& \lim_{T\rightarrow \infty}\frac{1}{TK}\sum_{t\in\mathcal{T}}\sum_{\tau\in\mathcal{T}_t} T_i^{tol}(\tau)\leq T_i^{max}/K,\forall i\in\mathcal{I}, \label{LongTermDelaySlot} \\
& \lim_{T\rightarrow \infty}\frac{1}{TK}\sum_{t\in\mathcal{T}}\sum_{\tau\in\mathcal{T}_t} E^{tol}(\tau)\leq E^{max}/K.\label{LongTermEnergySlot}
\end{align}
\end{subequations}
Note that the reformulated problem $\mathcal{P}_2$ is equivalent to the original problem $\mathcal{P}_1$ with exactly the same decision variables remaining in two different timescales, while all long-term constraints have been unified into a single timescale (i.e., in terms of the time slot only) but will not affect the optimization performance.

Obviously, $\mathcal{P}_2$ is still a long-term optimization problem, and the major difficulties for solving it are i) how to address the long-term average delay and energy consumption constraints; and ii) how to optimize two-timescale decision variables simultaneously. To this end, in the next subsection, we employ the Lyapunov method \cite{neely2022stochastic}, and reformulate $\mathcal{P}_2$ to accommodate the two-timescale features.

\subsection{Problem Decomposition}
We first define a delay overflow queue and an energy deficit queue to respectively describe how task response delay $T_i^{tol}(\tau)$ of each PT $i\in\mathcal{I}$ and the overall system energy consumption $E^{tol}(\tau)$ in time slot $\tau \in \mathcal{T}_t$ may deviate from the long-term budget $T_i^{max}/K$ and $E^{max}/K$.
The dynamic evolution of these two queues can be expressed as
\begin{equation} \label{DelayQueue}
H_i(\tau+1)=\max[H_i(\tau)+T_i^{tol}(\tau)-T_i^{max}/K,0],
\end{equation}
\begin{equation} \label{EnergyQueue}
E(\tau+1)=\max[E(\tau)+E^{tol}(\tau)-E^{max}/K,0].
\end{equation}

After that, we combine the delay overflow queue $H_i(\tau)$ for all tasks of PTs and energy deficit queue $E^{tol}(\tau)$ by a vector as $\boldsymbol\Theta(\tau)=[\boldsymbol{H}(\tau),E(\tau)]$, and introduce a quadratic Lyapunov function as \cite{neely2022stochastic}:
\begin{equation}
L(\boldsymbol\Theta(\tau))\triangleq\frac{1}{2}[\sum_{i\in\mathcal{I}}H_i(\tau)^2+E(\tau)^2],
\end{equation}
which quantitatively reflects the congestion of all queues, and should be persistently pushed towards a minimum value to keep queue stabilities.
Referring to \cite{georgiadis2006resource}, the conditional Lyapunov drift is given by
\begin{equation}\label{drift}
\Delta(\boldsymbol\Theta(\tau))=\mathbb{E}[L(\boldsymbol\Theta(\tau+K))-L(\boldsymbol\Theta(\tau))|\boldsymbol\Theta(\tau)],
\end{equation}
where $\mathbb{E}[\cdot]$ denotes the expectation, and $\Delta(\boldsymbol\Theta(\tau))$ measures the difference of the Lyapunov function between $K$ consecutive time slots. Intuitively, by minimizing the Lyapunov drift in (\ref{drift}), we can prevent queue backlogs from the unbounded growth, and thus guarantee that the desired delay and energy consumption constraints can be met.

Accordingly, the Lyapunov drift-plus-penalty function becomes
\begin{equation}
\Delta(\boldsymbol{\Theta}(\tau))-V  \mathbb{E}[\sum_{i\in\mathcal I} A_i(\tau) \mid \boldsymbol{\Theta}(\tau)],
\end{equation}
where a control parameter $V > 0$ is introduced, representing the weight of significance on maximizing the HDT-assisted task execution accuracy versus that of strictly satisfying the delay and energy consumption constraints.
\begin{theorem} \label{Theorem1}
 Let $V > 0$, and the drift-plus-penalty is bounded by any possible decisions in any time slot $\tau\in\mathcal{T}_t$, i.e.,
 \begin{equation}\label{Theorem1Equ}
   \begin{aligned}& \Delta(\boldsymbol\Theta(\tau))-V \mathbb{E}[\sum_{i\in\mathcal I} A_i(\tau) \mid \boldsymbol{\Theta}(\tau)] \\
   & \leq  G+\sum_{i\in\mathcal{I}} \mathbb E[H_i(\tau) (T_i^{tol}(\tau)-T_i^{max}/K) \mid \boldsymbol\Theta(\tau)] \\
   &+ \mathbb E[E(\tau) (E^{tol}(\tau)-E^{max}/K) \mid \boldsymbol\Theta(\tau)] \\
   &-V   \mathbb E[\sum_{i \in \mathcal{I}} A_i(\tau) \mid \boldsymbol\Theta(\tau)],
   \end{aligned}
 \end{equation}
 where
 \begin{align}
   G\hspace{-1pt}=\hspace{-1pt}\frac{1}{2} \hspace{-1pt}\sum_{i\in\mathcal{I}}[T_i^{tol}\hspace{-1pt}(max)\hspace{-1pt}-\hspace{-1pt}T_i^{max}]^2\hspace{-1pt}+\hspace{-1pt}\frac{1}{2}[E^{tol}(max )\hspace{-1pt}-\hspace{-1pt}E^{max}]^2\hspace{-1pt}\nonumber
 \end{align}
is a positive constant that adjusts the tradeoff between the HDT-assisted task execution accuracy and the satisfaction degree of the delay and energy consumption constraints.
\end{theorem}
\begin{IEEEproof}
Please see Appendix \ref{AppendixA}.
\end{IEEEproof}

Theorem \ref{Theorem1} shows that the drift-plus-penalty is deterministically upper bounded in each time slot $\tau\in\mathcal{T}_t$.
Then, taking the sum over all time slots within time frame $t$ for both sides of (\ref{Theorem1Equ}), we have
\begin{align}\label{Theorem2Equ}
& \sum_{\tau\in\mathcal{T}_t}\Delta(\boldsymbol\Theta(\tau))-V \sum_{\tau\in\mathcal{T}_t}\mathbb{E}[\sum_{i\in \mathcal{I}} A_i(\tau) \mid \boldsymbol{\Theta}(\tau)] \nonumber\\
& \leq GK+\sum_{\tau\in\mathcal{T}_t} \sum_{i\in\mathcal{I}} \mathbb E[H_i(\tau) (T_i^{tol}(\tau)-T_i^{max}/K) \mid \boldsymbol\Theta(\tau)] \nonumber\\
&+\sum_{\tau\in\mathcal{T}_t} \mathbb E[E(\tau) (E^{tol}(\tau)-E^{max}/K) \mid \boldsymbol\Theta(\tau)] \nonumber\\
&-V \sum_{\tau\in\mathcal{T}_t} \mathbb E[\sum_{i\in\mathcal{I}} A_i(\tau) \mid \boldsymbol\Theta(\tau)].
\end{align}

Then, $\mathcal{P}_2$ can be decomposed into multiple subproblems, each of which opportunistically minimize the right-hand-side of (\ref{Theorem2Equ}) in one time frame $t \in \mathcal{T}$, i.e.,
\begin{align}
&\mathcal{P}_3:\min_{\mathcal J_i^A(t),\mathcal J_i^B(\tau)}\sum_{\tau \in \mathcal{T}_t}\sum_{i\in\mathcal{I}} \mathbb E[H_i(\tau)(T_i^{tol}(\tau) \nonumber\\
& -T_i^{max}/K)\mid \boldsymbol\Theta(\tau)]+\sum_{\tau \in \mathcal{T}_t} \mathbb E[E(\tau) (E^{tol}(\tau) \nonumber\\
& -E^{max}/K) \mid \boldsymbol\Theta(\tau)]-V \sum_{\tau \in \mathcal{T}_t} \mathbb E[\sum_{i\in\mathcal{I}} A_i(\tau) \mid \boldsymbol\Theta(\tau)] \nonumber\\
& \text { s.t. } (\ref{C1}),(\ref{C3}),(\ref{C4}).\nonumber
\end{align}

Note that, in problem $\mathcal{P}_3$, decisions $\mathcal J_i^A(t)= \{a_{i,m}(t),x_{i}(t) \}$ and $\mathcal J_i^B(\tau)=\{b_{i}(\tau),y_{i}^{}(\tau), f_{i}^{}(\tau),z_i(\tau)\}$ remain unchanged as those in $\mathcal{P}_3$. This means that, although $\mathcal{P}_3$ focuses on the optimization in a single time frame, it still includes two-timescale variables.

\subsection{Alternating Algorithm between Two Timescales}\label{Alternating}
\subsubsection{Two-Timescale Decoupling and Alternation}
To solve problem $\mathcal{P}_3$, we can decouple it into two subproblems (one for the large-timescale, and the other for the small-timescale), and then solve them alternately till the convergence.

\textbf{Large-timescale Problem:} Given the small-timescale decision $\mathcal J_i^B(\tau)$ together with the current backlogs of delay overflow queues $H_i(\tau)$ of all PTs and the system energy consumption deficit queue $E(\tau)$, the large-timescale subproblem aims to jointly optimize the granularity of experiential knowledge $x_i(t)$ and access selection $a_{i,m}(t)$ at the beginning of each time frame $t\in\mathcal{T}$, which can be formulated as
\begin{align}& \mathcal{P}_4:\min _{\mathcal J_i^A(t)} \sum_{i \in \mathcal{I}}  H_i(\tau)[(T_{i}^{dl}(t)+\sum_{m \in\mathcal M} T_{i, m}^{pl}(t))/K \nonumber\\
& + z_i(\tau)\sum_{m\in \mathcal M} (T_{i,m}^{ul}(\tau)+T_{i,m}^{ud}(\tau)+T_{i,m}^{ofld}(\tau)) \nonumber\\
& +T_{i}^{exec}(\tau)]+E(\tau)\sum_{i \in \mathcal{I}}[(E_{i}^{dl}(t)+\sum_{m \in\mathcal M}E_{i, m}^{pl}(t))/K \nonumber\\
& + z_i(\tau)\sum_{m \in\mathcal M}(E_{i,m}^{ul}(\tau)+E_{i, m}^{ud}(\tau)+E_{i, m}^{ofld}(\tau))   \nonumber\\
& +E_{i}^{exec}(\tau)]-V \sum_{i \in \mathcal{I}}A_i(\tau) \nonumber\\
& \text { s.t. } (\ref{C1}),(\ref{C3}),(\ref{C4}).\nonumber
\end{align}

\textbf{Small-timescale Problem:} Given the large-timescale decision $\mathcal J_i^A(t)$, the small-timescale subproblem targets to jointly optimize the personalized data size $y_i(\tau)$, bandwidth resource allocation $b_i(\tau)$, computation resource allocation $f_i(\tau)$ and task offloading $z_i(\tau)$ in each time slot $\tau\in\mathcal{T}_t$, which can be formulated as
\begin{align}
& \mathcal{P}_{5}:\min _{\mathcal J_i^B(\tau)} \sum_{i \in \mathcal{I}}  H_i(\tau)[\sum_{m\in\mathcal M}T_{i, m}^{pl}(t)/K+z_i(\tau)\sum_{m\in \mathcal M}  \nonumber\\
& (T_{i,m}^{ul}(\tau)+T_{i,m}^{ud}(\tau)+T_{i,m}^{ofld}(\tau))+T_{i}^{exec}(\tau)] \nonumber\\
& +E(\tau)\sum_{i \in \mathcal{I}} [\sum_{m \in\mathcal M}E_{i, m}^{pl}(t)/K+\sum_{m \in\mathcal M}z_i(\tau)(E_{i,m}^{ul}(\tau) \nonumber\\
& +E_{i, m}^{ud}(\tau)+E_{i, m}^{ofld}(\tau))+E_{i}^{exec}(\tau)]-V \sum_{i \in \mathcal{I}}A_i(\tau) \nonumber\\
& \text{ s.t. } (\ref{C3}),(\ref{C4}).\nonumber
\end{align}

\textbf{Alternating Process:} For $\tau=tK$ (i.e., the beginning of each time frame), we iteratively optimize subproblems $\mathcal{P}_{4}$ and $\mathcal{P}_{5}$ until the objective of $\mathcal{P}_{3}$ converges.
Specifically, by fixing small-timescale decisions as $\mathcal{J}_i^B(tK)=J_i^B(tK-1)$ (inherited from the last time slot $\tau=tK-1$ of the previous time frame $\mathcal{T}_{t-1}$), large-timescale subproblem $\mathcal{P}_4$ is first solved to optimize $\mathcal{J}_i^A(tK)$. Then, given large-timescale decisions $\mathcal{J}_i^A(tK)$, small-timescale subproblem $\mathcal{P}_5$ is solved to update $\mathcal{J}_i^B(tK)$, which is returned back to $\mathcal{P}_4$.
For each $\tau\in[tK+1,tK+K-1]$ (i.e., the rest of each time frame), with the optimized large-timescale decisions $\mathcal{J}_i^A(tK)$ after the convergence, we repeatedly solve $\mathcal{P}_5$ to obtain $\mathcal{J}_i^B(\tau)$ in each time slot.

\begin{algorithm}[!t]\label{CRBRA}
\footnotesize
\caption{PME-Based Algorithm } \label{algorithm2}
\KwIn{$N$ partitions, and initial feasible solution $z^{'}$}
Linear relaxation:  $a_{i,m}(t) \in\{0,1\} \rightarrow \tilde a_{i,m}(t) \in[0,1]$;\\
\For{$x_{i}^{}(t) \in \boldsymbol{x}$}
{
    \For{$\tilde a_{i,m}^{}(t) \in \boldsymbol{\tilde a}$}
    {
        \For{$n\in N$}
        {
            $x_i^{L}(t)=x_{i,n}^{L}(t);x_i^{U}(t)=x_{i,n}^{U}(t)$\\
            \If{(\ref{BC}) is feasible}
            {
                Tighten bound $[\tilde a_{i,m,n}^{L}(t),\tilde a_{i,m,n}^{U}(t)]$ from (\ref{BC});\\
            }
            \Else
            {
                Remove partition $[x_{i,n}^{L}(t),x_{i,n}^{U}(t)]$;\\
            }
        }
        Update lower bound as $\tilde a_{i,m}^{L}(t)=\min_n \tilde a_{i,m,n}^{L}(t)$;\\
        Update upper bound as $\tilde a_{i,m}^{U}(t)=\max_n \tilde a_{i,m,n}^{U}(t)$;\\
    }
    Update lower bound as $x_{i}^{L}(t)=\min_n x_{i,n}^{L}(t)$;\\
    Update upper bound as $x_{i}^{U}(t)=\max_n x_{i,n}^{U}(t)$;\\
}
Solve $\mathcal{P}_{4-3}$ in the pruned partition to get solution $(\boldsymbol{x}^R, \boldsymbol{\tilde a}^R)$;\\
Round $a_{i,m^*}(t)=\max_{m^*}\tilde a_{i,m}(t),\forall m\in\mathcal{M}$ to 1 and set the others to 0;\\
Get solution ($\boldsymbol{x}^R, \boldsymbol{a}^R$) of $\mathcal{P}_4$;\\
\KwOut{($\boldsymbol{x}^R, \boldsymbol{a}^R$)}
\end{algorithm}

\subsubsection{Solution for Large-Timescale Decisions} \label{Large}
Large-timescale subproblem $\mathcal{P}_4$ is non-convex in general, but can be regarded as a bilinear optimization problem (i.e., the problem is linear if we fix one decision variable and optimize the other one within $\mathcal{J}_i^A(t)$).
This motivates us to design a PME-based algorithm \cite{castro2015tightening}, which first constructs convex envelops for bilinear terms, transforming the problem to a piecewise linear form, and then solves it by \textit{partitioning} and \textit{pruning}.

Let $u_{i,m}(t)=a_{i,m}(t)x_{i}(t)$ be an auxiliary variable of bivariate $a_{i,m}(t)x_{i}(t)$. Then, $\mathcal{P}_4$ can be relaxed into a convex optimization problem as
\begin{subequations}
\begin{align}
& \mathcal{P}_{4-1}:\min _{\mathcal J_i^A(t), u_{i,m}(t)} \sum_{i \in \mathcal{I}}  H_i(\tau)[(T_{i}^{dl}(t)+\sum_{m \in\mathcal M} u_{i, m}(t) \nonumber\\
& \frac{D_i(t)C_m}{f_i(tK)F_m})/K+ z_i(\tau)\sum_{m\in \mathcal{M}} (T_{i, m}^{ul}(\tau)+T_{i, m}^{ud}(\tau)+T_{i, m}^{ofld}(\tau)) \nonumber\\
& +T_{i, m}^{exec}(\tau)]+E(\tau)\sum_{i \in \mathcal{I}} [(E_{i}^{dl}(t)+\sum_{m \in\mathcal M}u_{i,m}(t) \nonumber\\
& \rho_m(F_m)^2D_i(t)C_m)/K+z_i(\tau) \sum_{m \in \mathcal{M}}(E_{i, m}^{ul}(\tau) \nonumber\\
& +E_{i, m}^{ud}(\tau)+E_{i, m}^{ofld}(\tau))+E_{i}^{exec}(\tau)]-V  \sum_{i \in \mathcal{I}}A_i(\tau)  \nonumber\\
& \text { s.t. } (\ref{C1}),(\ref{C3}),(\ref{C4}), \nonumber\\
& u_{i, m}(t)\hspace{-2pt} \geq\hspace{-2pt} 0,\forall i\in\mathcal{I},\forall m\in\mathcal{M},\forall t\in\mathcal{T},    \label{C5}\\
& u_{i, m}(t)\hspace{-2pt} \geq \hspace{-2pt} x_{i}(t)\hspace{-2pt}+\hspace{-2pt}a_{i, m}(t)\hspace{-2pt}-\hspace{-2pt}1,\forall i\hspace{-2pt}\in\hspace{-2pt}\mathcal{I},\forall m\hspace{-2pt}\in\hspace{-2pt}\mathcal{M},\forall t\hspace{-2pt}\in\hspace{-2pt}\mathcal{T},  \label{C6} \\
& u_{i, m}(t)\hspace{-2pt} \leq \hspace{-2pt}a_{i, m}(t),\forall i\in\mathcal{I},\forall m\in\mathcal{M},\forall t\in\mathcal{T}, \label{C7} \\
& u_{i, m}(t)\hspace{-2pt} \leq\hspace{-2pt} x_{i}(t),\forall i\in\mathcal{I},\forall m\in\mathcal{M},\forall t\in\mathcal{T},    \label{C8}
\end{align}
\end{subequations}
where constraints (\ref{C5})-(\ref{C8}) are the corresponding relaxed ones on bivariate $a_{i,m}(t)x_{i}(t)$.

Next, to improve the solution quality, we divide $\boldsymbol{x} = \{x_{i}(t),\forall i\in\mathcal{I},\forall t\in\mathcal{T}\}$ and $\boldsymbol{a} = \{a_{i,m}(t),\forall i\in\mathcal{I},\forall i\in\mathcal{I},\forall t\in\mathcal{T}\}$ into $\mid\mathcal{N}\mid=N$ partitions.
Specifically, let $x_{i,n}(t)\in[x_{i,n}^L(t),x_{i,n}^U(t)]$ be the range of $x_{i,n}(t)$ in partition $n\in \mathcal{N}$, where $x_{i,n}^L(t)$ and $x_{i,n}^U(t)$ are its lower and upper bounds, respectively.
Besides, a new auxiliary variable $y_{i,n}(t)\in\{0,1\}$ is introduced, where $y_{i,n}(t)=1$ if the value of $x_{i}(t)$ belongs to partition $n$, and $y_{i,n}(t)=0$ otherwise.
Similarly, the binary variable $a_{i,m}(t)$ is first relaxed into a continuous variable $\tilde a_{i,m}(t)\in [0,1]$ and then divided into $N$ piecewise areas, where the range of partition $n\in\mathcal{N}$ is $\tilde  a_{i,m,n}(t)\in [\tilde a_{i,m,n}^L(t), \tilde a_{i,m,n}^U(t)]$.
Based on these, $\mathcal{P}_{4-1}$ can be converted into a generalized disjunctive programming problem as
\begin{align}
&\mathcal{P}_{4-2}:\min _{\mathcal J_i^A(t), u_{i,m}(t)} \sum_{i \in \mathcal{I}}  H_i(\tau)[(T_{i}^{dl}(t)+\sum_{m \in M} u_{i, m}(t) \nonumber\\
& \frac{D_i(t)C_m}{f_i(tK)F_m})/K+ z_i(\tau)\sum_{m\in \mathcal{M}} (T_{i, m}^{ul}(\tau)+T_{i, m}^{ud}(\tau)+T_{i, m}^{ofld}(\tau)) \nonumber\\
& +T_{i, m}^{exec}(\tau)]+E(\tau) \sum_{i \in \mathcal{I}}[(E_{i}^{dl}(t)+\sum_{m \in\mathcal M}u_{i,m}(t)\nonumber\\
& \rho_m(F_m)^2 D_i(t)C_m)/K+z_i(\tau) \sum_{m \in \mathcal{M}}(E_{i, m}^{ul}(\tau)+E_{i, m}^{ud}(\tau)\nonumber\\
& +E_{i, m}^{ofld}(\tau))+E_{i}^{exec}(\tau)]-V  \sum_{i \in \mathcal{I}}A_i(\tau) \nonumber\\
& \text { s.t. } (\ref{C1}), (\ref{C3}), (\ref{C4}), \nonumber\\
&\bigvee_{n\in\mathcal{N}}\left[\begin{array}{l}
u_{i, m}(t) \geqslant \tilde a_{i, m}(t) \cdot x_{i, n}^L(t)+\tilde a_{i, m, n}^L(t) \cdot x_i(t)\\-\tilde a_{i, m, n}^L(t) \cdot x_{i, n}^L(t) \\
u_{i, m}(t) \geqslant \tilde a_{i, m}(t) \cdot x_{i, n}^U(t)+\tilde a_{i, m, n}^U(t) \cdot x_i(t)\\-\tilde a_{i, m, n}^U(t) \cdot x_{i, n}^U(t) \\
u_{i, m}(t) \leqslant \tilde a_{i, m}(t) \cdot x_{i, n}^L(t)+\tilde a_{i, m, n}^U(t) \cdot x_i(t)\\-\tilde a_{i, m, n}^U(t) \cdot x_{i, n}^L(t) \\
u_{i, m}(t) \leqslant \tilde a_{i, m}(t) \cdot x_{i, n}^U(t)+\tilde a_{i, m, n}^L(t) \cdot x_i(t)\\-\tilde a_{i, m, n}^L(t) \cdot x_{i, n}^U(t) \\
\tilde a_{i, m, n}^L(t) \leqslant \tilde a_{i, m}(t) \leqslant \tilde a_{i, m, n}^U(t) \\
x_{i, n}^L(t) \leqslant x_i(t) \leqslant x_{i, n}^U(t) \\
y_{i,n}(t)
\end{array}\right],\nonumber\\
& x_{i, n}^L(t)=x_i^L(t)+\frac{(x_i^U(t)-x_i^L(t)) \cdot(n-1)}{N}, \forall i, \forall n, \forall t, \nonumber\\
& x_{i, n}^U(t)=x_i^L(t)+\frac{(x_i^U(t)-x_i^L(t)) \cdot n}{N}, \forall i, \forall n, \forall t, \nonumber\\
& y_{i,n}(t) \in\{0,1\}, \forall n, \forall t.\nonumber
\end{align}

Then, by applying the convex hull relaxation \cite{karuppiah2006global}, \cite{meyer2006global}, we can further transform problem $\mathcal{P}_{4-2}$ into a piecewise linear form as
\begin{align} \label{PRT-MILP}
&\mathcal{P}_{4-3}:\min _{\mathcal J_i^A(t), u_{i,m}(t)} \sum_{i \in \mathcal{I}}  H_i(\tau)[(T_{i}^{dl}(t)+\sum_{m \in M} u_{i, m}(t) \nonumber\\
& \frac{D_i(t)C_m}{f_i(tK)F_m})/K+ z_i(\tau)\sum_{m\in \mathcal{M}} (T_{i, m}^{ul}(\tau)+T_{i, m}^{ud}(\tau)+T_{i, m}^{ofld}(\tau)) \nonumber\\
& +T_{i, m}^{exec}(\tau)]+E(\tau) \sum_{i \in \mathcal{I}}[(E_{i}^{dl}(t)+\sum_{m \in\mathcal M}u_{i,m}(t)\nonumber\\
& \rho_m(F_m)^2 D_i(t)C_m)/K+z_i(\tau) \sum_{m \in \mathcal{M}}(E_{i, m}^{ul}(\tau)+E_{i, m}^{ud}(\tau)\nonumber\\
& +E_{i, m}^{ofld}(\tau))+E_{i}^{exec}(\tau)]-V  \sum_{i \in \mathcal{I}}A_i(\tau) \nonumber\\
& \text { s.t. } (\ref{C1}),(\ref{C3}),(\ref{C4}), \nonumber\\
& u_{i, m}(t) \geqslant \hat{a}_{i, m, n}(t) \cdot x_{i, n}^L(t)+\tilde a_{i, m, n}^L(t) \cdot \hat{x}_{i, n}(t)\nonumber\\&-\tilde a_{i, m, n}^L(t) \cdot x_{i, n}^L(t) \cdot y_{i,n}(t), \forall i, \forall m, \forall n, \forall t, \nonumber\\
& u_{i, m}(t) \geqslant \hat{a}_{i, m, n}(t) \cdot x_{i, n}^u(t)+\tilde a_{i, m, n}^U(t) \cdot \hat{x}_{i, n}(t)\nonumber\\&-\tilde a_{i, m, n}^U(t) \cdot x_{i, n}^u(t) \cdot y_{i,n}(t), \forall i, \forall m, \forall n, \forall t, \nonumber\\
& u_{i, m}(t) \leqslant \hat{a}_{i, m, n}(t) \cdot x_{i, n}^L(t)+\tilde a_{i, m, n}^U(t) \cdot \hat{x}_{i, n}(t)\nonumber\\&-\tilde a_{i, m, n}^U(t) \cdot x_{i, n}^L(t) \cdot y_{i,n}(t), \forall i, \forall m, \forall n, \forall t, \nonumber\\
& u_{i, m}(t) \leqslant \hat{a}_{i, m, n}(t) \cdot x_{i, n}^u(t)+\tilde a_{i, m, n}^L(t) \cdot \hat{x}_{i, n}(t)\nonumber\\&-\tilde a_{i, m, n}^L(t) \cdot x_{i, n}^u(t) \cdot y_{i,n}(t), \forall i, \forall m, \forall n, \forall t, \nonumber\\
& \tilde a_{i, m}(t)=\sum_{n\in \mathcal{N}} \hat{a}_{i, m, n}(t), \forall i, \forall m, \forall t,  \nonumber\\
& x_i(t)=\sum_{n\in \mathcal{N}} \hat{x}_{i, n}(t), \forall i, \forall t,  \nonumber\\
& \sum_{n\in\mathcal{N}} y_{i,n}(t)=1, \forall t, \nonumber\\
& x_{i, n}^L(t)=x_i^L(t)+\frac{(x_i^U(t)-x_i^L(t)) \cdot n(n-1)}{N}, \forall i, \forall n, \forall t,  \nonumber\\
& x_{i, n}^U(t)=x_i^L(t)+\frac{(x_i^U(t)-x_i^L(t)) \cdot n n}{N}, \forall i, \forall n, \forall t,  \nonumber\\
& x_{i, n}^L(t) y_{i,n}(t) \leqslant \hat{x}_{i, n}(t) \leqslant x_{i, n}^U(t)  y_{i,n}(t), \forall i, \forall n, \forall t,  \nonumber\\
& \tilde a_{i, m, n}^L(t) y_{i,n}(t) \leqslant \hat{a}_{i, m, n}(t) \leqslant \tilde a_{i, m, n}^U(t) y_{i,n}(t), \forall i, \forall m, \forall n, \forall t,  \nonumber\\
& y_{i,n}(t) \in\{0,1\}, \forall n, \forall t.\nonumber
\end{align}

To this end, we prune $x_i(t)$ and $\tilde a_{i,m}(t)$ to tighten their relaxed bounds.
By traversing each partition $n\in N$ of $x_{i}(t)$, we first determine the lower bound $\tilde a_{i,m,n}^L(t)$ and upper bound $\tilde a_{i,m,n}^U(t)$ of $\tilde a_{i, m}(t)$ by solving the following linear programming problem:
\begin{equation}\label{BC}
\begin{aligned}
& \tilde a_{i, m, n}^{L}(t)=\min\tilde a_{i, m}(t) \text { or }\tilde a_{i, m, n}^{U}(t)=\max\tilde a_{i, m}(t) \\
& \text { s.t. }(\ref{C1}), (\ref{C3}), (\ref{C4}), (\ref{C5})-(\ref{C8}), \\
&  obj(\mathcal{P}_{4-3}) \leqslant z^{\prime}, \\
&  x_i^{L}(t)=x_{i, n}^{L}(t) \leqslant x_i(t) \leqslant x_{i, n}^{U}(t)=x_i^{U}(t).
\end{aligned}
\end{equation}
Besides, the range of $x_{i}(t)$ is updated as $x_{i}^{L}(t)=\min_n x_{i,n}^{L}(t)$ and $x_{i}^{U}(t)=\max_n x_{i,n}^{U}(t)$ after traversing all the partitions of $x_{i}(t)$.
Then, after pruning all $x_i(t)$ and $\tilde a_{i,m}(t)$, we can solve problem $\mathcal{P}_{4-3}$ in the pruned partition with software-based optimization solvers (e.g., CVX \cite{bliek1u2014solving}), and obtain its solution $(\boldsymbol{x}^R, \boldsymbol{\tilde{a}}^R)$.
Lastly, we round continuous variables $\boldsymbol{\tilde{a}}^R$ to binary forms for obtaining integer solutions.
Note that all constraints are automatically satisfied under $(\boldsymbol{x}^R, \boldsymbol{a}^R)$ because they have been taken into account in the pruning process of (\ref{BC}).
The detailed steps of the designed PME-based algorithm for solving the large-timescale problem is presented in Algorithm \ref{algorithm2}.


\subsubsection{Solution for Small-Timescale Decisions}\label{Small}
Small-timescale subproblem $\mathcal{P}_5$ is also non-convex in general, but by relaxing the integer decision $z_i(\tau)$ (i.e., $z_i(\tau)\in\{0,1\}\rightarrow z_i(\tau)\in[0,1]$), it becomes a block multi-convex problem (i.e., the problem is convex if we solve one block of decision variable while fixing the others).
This motivates us to design a BCD-based algorithm by further dividing problem $\mathcal{P}_{5}$ into four subproblems and solving them alternately until the objective function of $\mathcal{P}_{5}$ converges.

\textit{Personalized Data Size Determination:} Given $b_{i}(\tau)$, $f_{i}^{}(\tau)$ and $z_i(\tau)$, we have
\begin{align}
& \mathcal{P}_{5-1}: \min _{y_i(\tau)} \sum_{i \in \mathcal{I}}  H_i(\tau)z_i(\tau)\sum_{m\in\mathcal M} (T_{i,m}^{ul}(\tau)+T_{i,m}^{ud}(\tau)) \nonumber\\
& +E(\tau)\sum_{i \in \mathcal{I}} \sum_{m \in \mathcal M}z_i(\tau)(E_{i,m}^{ul}(\tau)+E_{i, m}^{ud}(\tau))-V \sum_{i \in \mathcal{I}}A_i(\tau)\nonumber\\
& \text { s.t. } y_i(\tau)\in[0,1].\nonumber
\end{align}

\textit{Bandwidth Resource Allocation:} Given $y_{i}(\tau)$, $f_{i}^{}(\tau)$ and $z_i(\tau)$, we have
\begin{align}
& \mathcal{P}_{5-2}: \min_{b_i(\tau)} \sum_{i \in \mathcal{I}}  H_i(\tau)z_i(\tau)\sum_{m\in\mathcal M} (T_{i,m}^{ul}(\tau)+T_{i,m}^{ofld}(\tau)) \nonumber\\
& +E(\tau)\sum_{i \in \mathcal{I}} \sum_{m \in \mathcal M}z_i(\tau)(E_{i,m}^{ul}(\tau)+E_{i, m}^{ofld}(\tau))\nonumber\\
& \text { s.t. } (\ref{C3}), b_i(\tau)\in(0,1].\nonumber
\end{align}

\textit{Computation Resource Allocation:} Given $y_{i}(\tau)$, $b_{i}^{}(\tau)$ and $z_i(\tau)$, we have
\begin{align}
& \mathcal{P}_{5-3}: \min _{f_i(\tau)} \sum_{i \in \mathcal{I}}  H_i(\tau)(\sum_{m\in\mathcal M}T_{i, m}^{pl}(t)/K+z_i(\tau)\sum_{m\in \mathcal M}  \nonumber\\
& T_{i,m}^{ud}(\tau)+T_{i}^{exec}(\tau))+E(\tau)\sum_{i \in \mathcal{I}} (\sum_{m \in\mathcal M}E_{i, m}^{pl}(t)/K \nonumber\\
& +\sum_{m \in\mathcal M}z_i(\tau)E_{i, m}^{ud}(\tau)+E_{i}^{exec}(\tau)) \nonumber\\
& \text { s.t. } (\ref{C4}), f_i(\tau)\in(0,1].\nonumber
\end{align}

\begin{algorithm}[!t]
\footnotesize
\caption{BCD-Based Algorithm} \label{algorithm3}
\KwIn{Iteration index $l=0$, iterative convergence threshold $\epsilon$, and initial feasible solutions ($\boldsymbol{y}^{(0)},\boldsymbol{b}^{(0)},\boldsymbol{f}^{(0)},\boldsymbol{z}^{(0)}$).}
Linear relaxation:  $z_{i}(\tau) \in\{0,1\} \rightarrow z_{i}(\tau) \in[0,1]$;\\
\Repeat{$\mid obj^{(l)}(\mathcal{P}_5)-obj^{(l-1)}(\mathcal{P}_5) \mid\leq \epsilon$}
{
Solve $\boldsymbol y^{(l)}$ from $\mathcal{P}_{5-1}$ with given $\boldsymbol{b}^{(l-1)},\boldsymbol{f}^{(l-1)},\boldsymbol{z}^{(l-1)}$;\\
Solve $\boldsymbol b^{(l)}$ from $\mathcal{P}_{5-2}$ with given $\boldsymbol{y}^{(l-1)},\boldsymbol{f}^{(l-1)},\boldsymbol{z}^{(l-1)}$;\\
Solve $\boldsymbol f^{(l)}$ from $\mathcal{P}_{5-3}$ with given $\boldsymbol{y}^{(l-1)},\boldsymbol{b}^{(l-1)},\boldsymbol{z}^{(l-1)}$;\\
Solve $\boldsymbol z^{(l)}$ from $\mathcal{P}_{5-4}$ with given $\boldsymbol{y}^{(l-1)},\boldsymbol{b}^{(l-1)},\boldsymbol{f}^{(l-1)}$;\\
$l=l+1$;
}
Round $z_i(\tau)$ to 1 with probability $z_i(\tau)$;\\
\KwOut{ solution of $\mathcal{P}_{5}$: ($\boldsymbol y^{(l)}, \boldsymbol b^{(l)}, \boldsymbol f^{(l)}, \boldsymbol z^{(l)}$). }
\end{algorithm}

\textit{Task Offloading Decision:} Given $y_{i}(\tau)$, $b_{i}^{}(\tau)$ and $f_i(\tau)$, we have
\begin{align}
& \mathcal{P}_{5-4}: \min _{z_i(\tau)} \sum_{i \in \mathcal{I}}  H_i(\tau)[z_i(\tau)\sum_{m\in \mathcal M}(T_{i,m}^{ul}(\tau)+T_{i,m}^{ud}(\tau)  \nonumber\\
& +T_{i,m}^{ofld}(\tau))+T_{i}^{exec}(\tau)]+E(\tau)\sum_{i \in \mathcal{I}} [\sum_{m \in\mathcal M}z_i(\tau)(E_{i,m}^{ul}(\tau) \nonumber\\
& +E_{i, m}^{ud}(\tau)+E_{i, m}^{ofld}(\tau))+E_{i}^{exec}(\tau)]-V \sum_{i \in \mathcal{I}}A_i(\tau)\nonumber\\
& \text { s.t. } z_i(\tau)\in[0,1].\nonumber
\end{align}

\begin{theorem}\label{lemma4}
  Problems $\mathcal{P}_{5-1}$, $\mathcal{P}_{5-2}$, $\mathcal{P}_{5-3}$ and $\mathcal{P}_{5-4}$ are all convex.
\end{theorem}
\begin{IEEEproof}
 Please see Appendix B.
\end{IEEEproof}

Thanks to Theorem \ref{lemma4}, we can easily solve problems $\mathcal{P}_{5-1}$, $\mathcal{P}_{5-2}$, $\mathcal{P}_{5-3}$ and $\mathcal{P}_{5-4}$ by leveraging existing software-based optimization solvers (e.g. CVX \cite{bliek1u2014solving}).
Note that all these problems have to be solved iteratively, and the iteration terminates whenever the objective of problem $\mathcal{P}_5$ can no longer be enhanced. The detailed steps of the designed BCD-based algorithm for solving the small-timescale problem is illustrated in Algorithm \ref{algorithm3}.

\begin{figure}[!t]
\begin{center}
\includegraphics[width=9 cm]{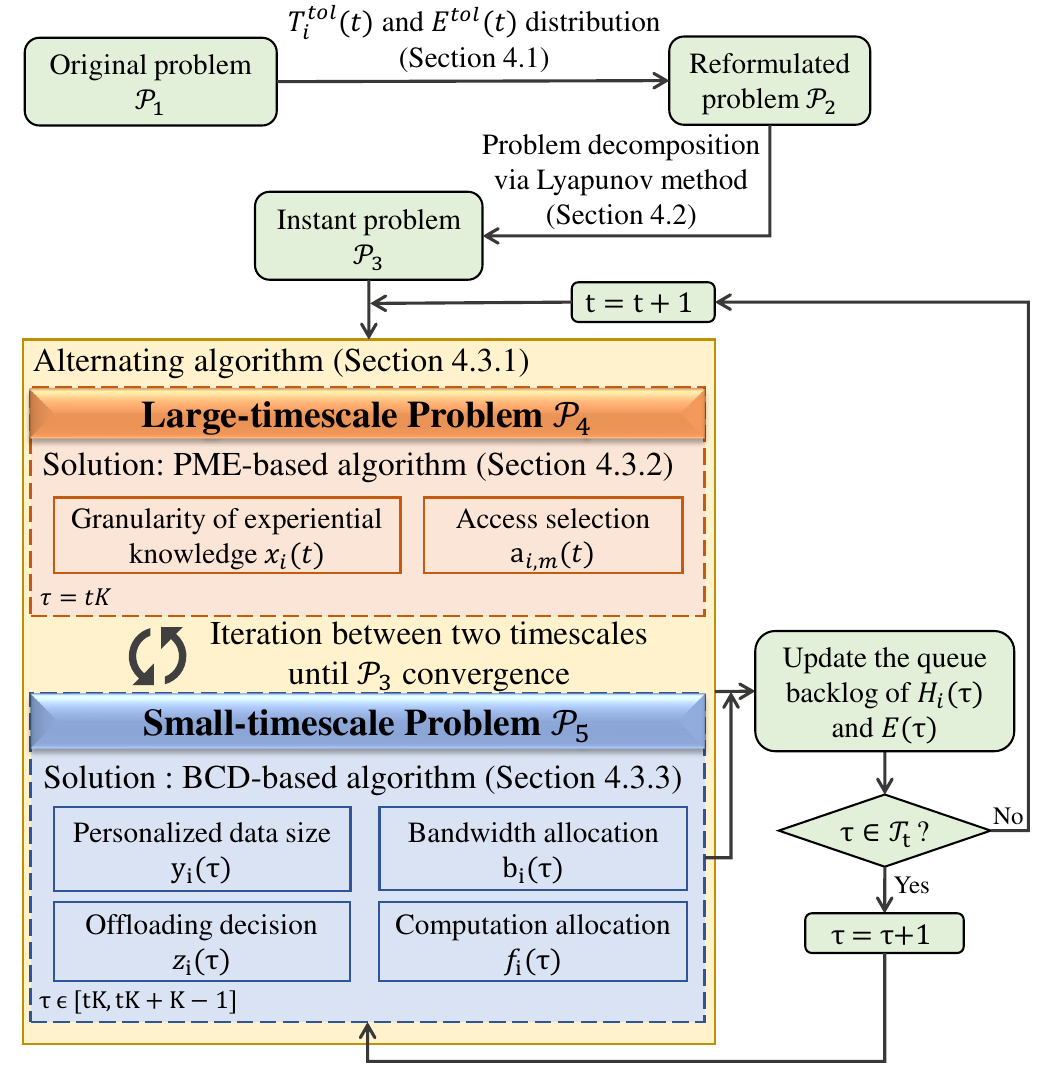}
  \caption{Flowchart of the proposed TACO approach.}
  \label{SolutionFramework}
\end{center}
\end{figure}

\subsection{Analysis of Proposed TACO Approach}\label{PA}

In summary, the proposed two-timescale accuracy-aware online optimization approach (TACO) consists of problem reformulation, decomposition and alternation between two timescales. The flowchart of TACO is shown in Fig. \ref{SolutionFramework}.


\begin{theorem} \label{TheoremConverge}
  The proposed TACO approach can converge with limited alternations and iterations.
\end{theorem}

\begin{IEEEproof}
For problem $\mathcal{P}_5$, the convergence of TACO depends on that of the designed BCD-based algorithm and that of the employed alternating algorithm for problem $\mathcal{P}_3$.

First, to prove the convergence of the BCD-based algorithm, we derive the partial derivatives of the objective function of problem $\mathcal{P}_5$ as follows:
\begin{align}\label{eq49}
& \nabla_{\boldsymbol{y}}obj(\mathcal{P}_5)= \sum_{i \in \mathcal{I}} H_i(\tau)z_i(\tau) \sum_{m \in\mathcal M}(\frac{S_i(\tau)}{r_{i, m}(\tau)} \\
& +\frac{a_{i, m}(t)S_i(\tau) C_m}{f_i(\tau) F_m}) +E(\tau)\sum_{i \in \mathcal{I}} z_i(\tau)\sum_{m \in\mathcal M}(\frac{S_i(\tau) p_i}{r_{i, m}(\tau)} \nonumber \\
& +a_{i, m}(t)\rho_m(F_m)^2 S_i(\tau) C_m)-2 V \sum_{i \in \mathcal{I}} \frac{z_i(\tau)S_i(\tau)}{D_i(t)+S_i(\tau)},\nonumber
\end{align}

\begin{align}\label{eq50}
& \nabla_{\boldsymbol{b}}obj(\mathcal{P}_5)=[\sum_{i\in\mathcal I} H_i(\tau) z_i(\tau) (y_i(\tau) S_i(\tau)\nonumber\\
& + \lambda_i(\tau))+E(\tau) \sum_{i\in\mathcal I}z_i(\tau)p_i (y_i(\tau) S_i(\tau)+\lambda_i(\tau))]  \nonumber\\
& [\frac{p_i|h_{i ,m}(\tau)|^2}{a_{i,m}(t)(S_{i,m}(\tau))^\theta N_0 (\tau)(B_m)^2 \ln 2}  \\
& \cdot \frac{(S_{i,m}(\tau))^\theta N_0(\tau)B_m}{((b_i(\tau))^3+(b_i(\tau))^2p_i|h_{i ,m}(\tau)|^2)} \nonumber\\
& \cdot \frac{1}{\log _2^2(1+\frac{p_i|h_{i ,m}(\tau)|^2}{(S_{i,m}(\tau))^\theta N_0 b_{i}(\tau)B_m})}\nonumber\\
& -\frac{1}{a_{i,m}(t)(b_i(\tau))^2 B_m \log _2(1+\frac{p_i|h_{i ,m}(\tau)|^2}{(S_{i,m}(\tau))^\theta N_0 b_{i}(\tau)B_m})}],\nonumber
\end{align}

\begin{equation}\label{eq51}
\begin{aligned}
& \nabla_{\boldsymbol{f}}obj(\mathcal{P}_5)= -\sum_{i \in \mathcal{I}}  H_i(\tau)[\sum_{m\in\mathcal M}\frac{a_{i,m}(t)x_{i}(t)D_i(t) C_m}{(f_{i}(tK))^2F_m K}\\
& +z_i(\tau)\sum_{m\in \mathcal M}\frac{a_{i,m}(t)y_{i}(\tau)S_i(\tau) C_m}{(f_{i}(\tau))^2F_m}\\
& +\sum_{m\in\mathcal{M}}a_{i,m}(t) z_i(\tau)\frac{\lambda_i(\tau) C_m}{(f_{i}(\tau))^2F_m}],
\end{aligned}
\end{equation}

\begin{align}\label{eq52}
& \nabla_{\boldsymbol{z}}obj(\mathcal{P}_5)=\sum_{i\in\mathcal I} H_i(\tau)[\sum_{m \in M} (T_{i, m}^{ul}(\tau)+T_{i, m}^{ud}(\tau)\nonumber\\
& +T_{i, m}^{ofld}(\tau))+\sum_{m\in\mathcal{M}}a_{i,m}(t) \frac{\lambda_i(\tau) C_m}{f_{i}(\tau)F_m}-\frac{\lambda_i(\tau)C_i}{F_i}] \nonumber\\
& +E(\tau)\sum_{i \in \mathcal I}[\sum_{m \in \mathcal M} (E_{i, m}^{ul}(\tau)+E_{i, m}^{ud}(\tau)+E_{i, m}^{ofld}(\tau))\\
& +\sum_{m\in\mathcal{M}}a_{i,m}(t)\rho_m(F_m)^2\lambda_i(\tau) C_m-\rho_i(F_i)^2\lambda_i(\tau)C_i]\nonumber\\
& -V\sum_{i\in\mathcal{I}}[1-(1-\frac{x_i(t)D_i(t)+y_i(\tau)S_i(\tau)}{D_i(t)+S_i(\tau)})^2-g_i^{local}].\nonumber
\end{align}
Evidently, (\ref{eq49}) and (\ref{eq52}) are constants, and (\ref{eq50}) and (\ref{eq51}) are linear, meaning that all derived partial derivatives are L-lipschitz continuous according to \cite{bolte2014proximal}, and hence the BCD-based algorithm can converge with limited iterations.

Then, to prove the convergence of the alternating algorithm, we scale the objective function of $\mathcal{P}_3$ as
\begin{align}\label{eq48}
& obj^{l-1}(\mathcal{P}_3) = obj(\mathcal{P}_3)(\boldsymbol{a}^{l-1},\boldsymbol{x}^{l-1},\boldsymbol{y}^{l-1},\boldsymbol{b}^{l-1},\boldsymbol{f}^{l-1},\boldsymbol{z}^{l-1})\nonumber\\
& \geq obj(\mathcal{P}_3)(\boldsymbol{a}^{l},\boldsymbol{x}^{l},\boldsymbol{y}^{l-1},\boldsymbol{b}^{l-1},\boldsymbol{f}^{l-1},\boldsymbol{z}^{l-1})\\
& \geq obj(\mathcal{P}_3)(\boldsymbol{a}^{l},\boldsymbol{x}^{l},\boldsymbol{y}^{l},\boldsymbol{b}^{l},\boldsymbol{f}^{l},\boldsymbol{z}^{l}) = obj^l(\mathcal{P}_3),\nonumber
\end{align}
where $l$ is the number of alternations.
The first inequality holds due to the sub-optimality of $\{\boldsymbol{y}^{l},\boldsymbol{b}^{l},\boldsymbol{f}^{l},\boldsymbol{z}^{l}\}$ by BCD-based algorithm,
and the second inequality holds because of the sub-optimality of $\{\boldsymbol{a}^{l},\boldsymbol{x}^{l}\}$ by PME-based algorithm. This indicates that the objective function of $\mathcal{P}_3$ is monotonically decreasing along with the alternating process, and will be lower-bounded by $-VKI$ (i.e., by setting two queue backlogs to $0$ and all $A_i(\tau)$ to $1$) in finite alternations.
\end{IEEEproof}


\begin{theorem}\label{TheoremComplexity}
  The computational complexity of the proposed TACO approach is $O(T R^{max}((I+M)log_2(I+M)+ I^3M^3N^3+IM^{2.055}+4W^{max}I^3+2^{2.055}I))$, where $I$ is the number of PTs, $M$ is the number of ESs, $N$ is the number of partitions in the PME-based algorithm, $T$ is the number of time frames, $R^{max}$ is the number of iterations in the alternating algorithm between two timescales, and $W^{max}$ is the number of iterations in the BCD-based algorithm.
\end{theorem}

\begin{IEEEproof}
The complexity of TACO mainly depends on the alternating algorithm integrating PME-based algorithm and BCD-based algorithm.

For the PME-based algorithm, as stated in \cite{peixoto2014efficient}, the computational complexity for obtaining an initial feasible solution is $O((I+M)log_2(I+M))$, and that of solving (\ref{BC}) is $O(I^3M^3N^3)$ with the interior point method \cite{boyd2004convex} in the CVX solver. Besides, the linear relaxation and linear programming for solving $a_{i,m}(t)$ has an asymptotic computational complexity of $O(IM^{2.055})$ \cite{srinivasan1999approximation}. 
For the BCD-based algorithm, the total computational complexity for solving all subproblems is $O(4I^3)$ with the interior point method \cite{boyd2004convex} in CVX solver, and that of the linear relaxation and linear programming solver for solving $z_i(\tau)$ is $O(2^{2.055}I)$.

To sum up, the computational complexity of TACO can be expressed as $O(T R^{max}((I+M)log_2(I+M)+ I^3M^3N^3+IM^{2.055}+4W^{max}I^3+2^{2.055}I))$.
\end{IEEEproof}


\begin{theorem}\label{Theorem4}
  Given Lyapunov control parameter $V$, the optimality gap between the solution obtained by the proposed TACO approach and the theoretically optimal solution to problem $\mathcal{P}_1$ can be expressed as
  \begin{equation}\label{optimal}
    \hspace{-5pt}\sum_{t\in\mathcal{T}}\hspace{-3pt}\sum_{\tau \in \mathcal{T}_t}\hspace{-2pt} \mathbb{E}[\sum_{i\in\mathcal I} \hspace{-2pt}A_i(\tau) \hspace{-3pt}\mid \hspace{-3pt}\boldsymbol{\Theta}(\tau)] / KT \hspace{-2pt}- \hspace{-1pt}\mathcal{O} \hspace{-2pt}\leq\hspace{-2pt} G \hspace{-1pt}/\hspace{-1pt} V \hspace{-2pt}+ \hspace{-2pt}(\Lambda \hspace{-2pt}+ \hspace{-2pt}\Gamma)\hspace{-2pt} /\hspace{-1pt} VT,
  \end{equation}
  where $\mathcal{O}$ stands for the theoretically optimal solution, $\Lambda$ is the optimality gap of the PME-based algorithm, $\Gamma$ represents the optimality gap of the BCD-based algorithm, and $G$ is defined in (\ref{Theorem1Equ}).
\end{theorem}

\begin{IEEEproof}
First, inequality (\ref{Theorem2Equ}) can be intuitively expanded and rewritten as
\begin{equation}\label{EqX}
\begin{aligned}
& \sum_{\tau\in\mathcal{T}_t}\Delta(\boldsymbol\Theta(\tau))-V \cdot \sum_{\tau\in\mathcal{T}_t} \mathbb{E}[\sum_{i\in\mathcal I} A_i(\tau) \mid \boldsymbol{\Theta}(\tau)]\\
& \leq GK + \sum_{\tau\in\mathcal{T}_t}\{\sum_{i\in\mathcal{I}} \mathbb E[H_i(\tau) [T_i^{tol}(\tau)-T_i^{max}] \mid \boldsymbol\Theta(\tau)] \\&+ \mathbb E\{E(\tau) \cdot[E^{t o l}(\tau)-E^{max}] \mid \boldsymbol\Theta(\tau)\} \\
& -V \cdot  \mathbb E[\sum_{i \in \mathcal{I}} A_i(\tau) \mid \boldsymbol\Theta(\tau)]\}\\
& \leq GK + V \cdot K\mathcal{O} + \Lambda + \Gamma.
\end{aligned}
\end{equation}

Then, by summing up (\ref{EqX}) over $T$ time frames, we have
\begin{equation}\label{EqY}
\begin{aligned}
& (G+V\cdot \mathcal{O}+\frac{1}{K}(\Lambda + \Gamma))\cdot KT\\
& \geq\sum_{t\in T} \{\sum_{\tau\in\mathcal{T}_t}\Delta(\boldsymbol\Theta(\tau))-V \cdot \sum_{\tau\in\mathcal{T}_t} \mathbb{E}[\sum_{i\in\mathcal I} A_i(\tau) \mid \boldsymbol{\Theta}(\tau)]\}\\
& =\mathbb{E}[L(\boldsymbol{\Theta}(KT))-L(\boldsymbol{\Theta}(0))] \\
& + V \cdot \sum_{t\in T} \sum_{\tau\in\mathcal{T}_t}\mathbb E[\sum_{i \in \mathcal{I}} A_i(\tau) \mid \boldsymbol\Theta(\tau)]\\
& =\mathbb{E}[L(\boldsymbol{\Theta}(KT))]-\mathbb{E}[L(\boldsymbol{\Theta}(0))] \\
& + V \cdot \sum_{t\in T} \sum_{\tau\in\mathcal{T}_t}\mathbb E[\sum_{i \in \mathcal{I}} A_i(\tau) \mid \boldsymbol\Theta(\tau)]
\end{aligned}
\end{equation}
Afterwards, by moving $\mathbb{E}[L(\boldsymbol{\Theta}(0))]$ to the left-hand-side of (\ref{EqY}), and then dividing both sides by $TV$, inequality (\ref{optimal}) can be obtained.

Next, we further analyze the optimality gaps $\Lambda$ and $\Gamma$ as follows. For $\Lambda$, according to \cite{castro2015tightening}, its value is theoretically bounded in the range of  $\Lambda\in[0, 0.12]$.
For $\Gamma$, by letting $\tilde {\mathcal{P}}_{5}$ and $\mathcal{P}_{5}^*$ be the solutions given by the BCD-based algorithm and the optimal one, respectively, and taking the subtraction between them, we have
\begin{align}
& \tilde{\mathcal{P}}_{5}-\mathcal{P}_{5}^*\nonumber\\
& \leq \Gamma   \nonumber\\
& =\sum_{i \in \mathcal I} H_i(\tau)\{\tilde{z}_i(\tau) \sum_{m \in \mathcal M}(S_i(\tau)+\lambda_i(\tau)+a_{i, m}(t) S_i(\tau) C_m) \nonumber\\
& +z_i^*(\tau) a_{i, m}(t) S_i(\tau) C_m\}+E(\tau) \sum_{i \in \mathcal I} \sum_{m \in \mathcal M}[\tilde{z}_i(\tau)  \nonumber\\
& [a_{i, m}(t) \rho_m(F_m)^2 S_i(\tau) C_m +S_i(\tau)p_i+z_i^*(\tau) S_i(\tau) p_i]  \nonumber\\
& +V \sum_{i \in \mathcal I}(A_i^*(\tau) \tilde{A}_i(\tau))=\sum_{i \in \mathcal I} H_i(\tau)(\tilde{z}_i(\tau)  \omega_i(\tau)+z_i^*(\tau) \phi_i(\tau))\nonumber\\
& +E(\tau) \sum_{i \in \mathcal I} \sum_{m \in \mathcal M}[\tilde{z}_i(\tau) \omega_i^{\prime}(\tau)+z_i^*(\tau) \phi_i^{\prime}(\tau)] \nonumber\\
& +V \sum_{i \in \mathcal I}(A_i^*(\tau)+\tilde{A}_i(\tau)),
\end{align}
where $\omega_{i}(\tau)=\sum_{m \in \mathcal M}(S_i(\tau)+\lambda_i(\tau)+a_{i, m}(t) S_i(\tau) C_m)$,
$\phi_i(\tau)=a_{i, m}(t) S_i(\tau) C_m$,
$\omega_{i}^{\prime}(\tau)=a_{i, m}(t) \rho_m(F_m)^2 S_i(\tau) C_m$
and $\phi_{i}^{\prime}(\tau)=S_i(\tau)p_i$, which are both constant in each time slot $\tau \in \mathcal{T}_t$.
\end{IEEEproof}

\section{Simulation Results}\label{PE}
In this section, simulations are conducted to evaluate the performance of the proposed TACO approach for jointly optimizing the HDT deployment (i.e., generic placement and customized update of the VT model) and task offloading in an end-edge-cloud collaborative framework. All simulation results are obtained based on real-world datasets (including a human activity dataset \cite{mi12:ubicomp-sagaware} and an ES distribution dataset \cite{telecom2019distribution}), by taking averages over 1000 runs under various parameter settings.

\subsection{Simulation Settings}
Consider an HDT system in a $1km \times 1km$ square area with $M=10$ ESs and $I=40$ PTs (following a Random-Waypoint (RWP) mobility model under the same settings as those in \cite{johnson1996dynamic}).
According to \cite{wu2020accuracy}, for the HDT-assisted complex task execution, the average accuracy of edge execution and local execution for any PT $i$' s tasks are approximated as $g_i^{edge}(d_i(\tau))=1-[1-d_i(\tau)/(D_i(t)+S_i(\tau))]^2$ (which is a function of the total data size for PT $i$'s corresponding VT construction at the edge, i.e., $d_i(\tau)$) and $g_i^{local}=0.5$, respectively.
Table \ref{ParaTable} lists the values of main simulation parameters, while most of them have also been widely employed in the literature \cite{wang2022online}, \cite{ouyang2018follow}, \cite{lin2021stochastic}. 
Furthermore, to show the superiority of the proposed TACO approach, the following schemes are simulated as benchmarks. Note that, since the original objectives of these benchmark schemes are different from ours, for the fairness of comparison, we have modified them to adapt to the considered settings and particularly changed their optimization objectives to align with ours.
\begin{itemize}

\item LOT \cite{lin2021stochastic}:    
Generic VT model placement from the cloud and PTs' task offloading are jointly determined by adopting a contract theory-based incentive mechanism. However, this scheme ignores the customized VT update by collecting personalized data from end devices, and it optimizes all decisions synchronously in a single-timescale only.

\item CRO \cite{10301793}:   
Both generic VT model placement from the cloud and customized VT update from end devices along with PTs' task offloading are jointly determined by adopting a double auction based optimization, while all decisions are optimized in a single-timescale for simplicity.



\end{itemize}

\begin{table}[!t]
\begin{center}
\footnotesize
\renewcommand{\arraystretch}{1.1}
\caption{Simulation Parameters}
\label{ParaTable}
\begin{tabular}{p{35pt}|p{60pt}|p{35pt}|p{65pt}}
\hline\noalign{\smallskip}
\textbf{Parameter} & \textbf{Value} & \textbf{Parameter} & \textbf{Value} \\
\noalign{\smallskip}\hline\noalign{\smallskip}
$p_i$               & 500 mW                 & $B_m$             & 5 MHz              \\
$p^c$               & 5 W                    & $F_m$             & 20 GHz             \\
$\theta$            & 4                      & $C_m,C_i$         & 300 cycles/bit     \\
$N_0$               & -174 dBm/Hz            & $\rho_m,\rho_i$   & $10^{-27}$         \\
$S_i(\tau)$         & [6.1, 12.2] Mbits      & $r^c$             & 50 Mbps            \\
$D_i(t)$            & [73.2, 97.6] Mbits     & $K$               & 10                 \\
$\lambda_i(\tau)$   & [10, 20] Mbits         & $M$               & 10                 \\
$F_i$               & 1 GHz                  & $I$               & 40                 \\
\hline
\end{tabular}
\end{center}
\end{table}


\subsection{Performance Evaluations}




\begin{figure}[!t]
\begin{center}
\includegraphics[width=7 cm]{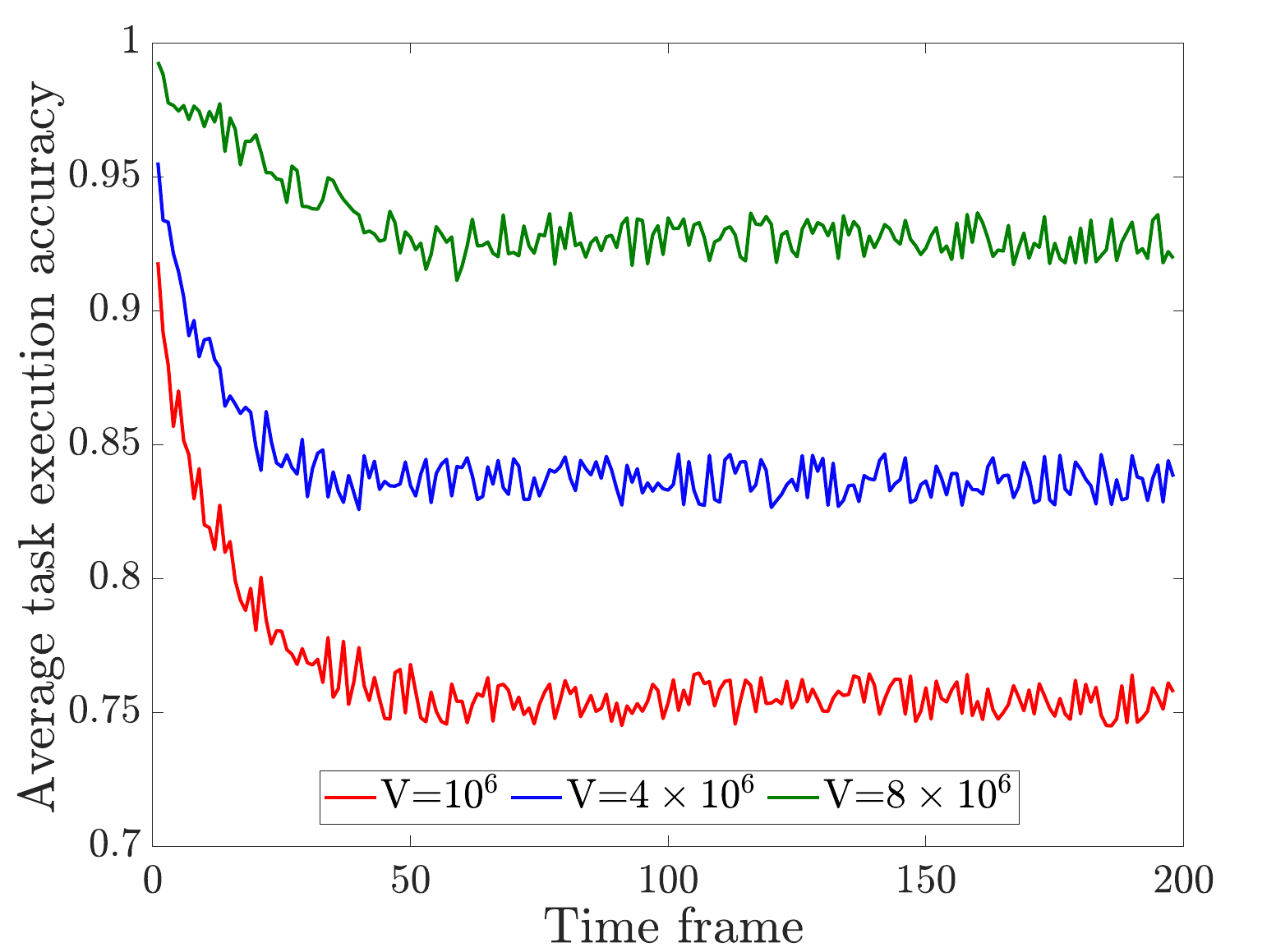}
  \caption{Convergence of the proposed TACO approach.}
  \label{AccuracyConverge}
\end{center}
\end{figure}

Fig. \ref{AccuracyConverge} examines the convergence of the proposed TACO approach in solving problem $\mathcal{P}_1$ by showing the average HDT-assisted task execution accuracy under different values of parameter $V$ ranging from $10^6$ to $8\times10^6$. The timeline is divided into $T=200$ time frames, and each frame has $K=10$ time slots.
It is shown that, for all three cases, within $50$ time frames, the task execution accuracy decreases at the beginning but quickly converges over the time, which verifies the convergence property of TACO in solving problem $\mathcal{P}_1$.
Besides, from this figure, it can also be observed that the average task execution accuracy increases with $V$. The reason is that, Lyapunov parameter $V$ is introduced to control the weight of significance on maximizing the HDT-assisted task execution accuracy versus that of enforcing the delay and energy consumption constraints, and a larger $V$ indicates more emphasis on the task execution accuracy.

\begin{figure}[!t]
\begin{center}
\hspace{-1.8em}
\subfigure[Queue backlogs w.r.t $V$.]{
\includegraphics[width=4.8 cm]{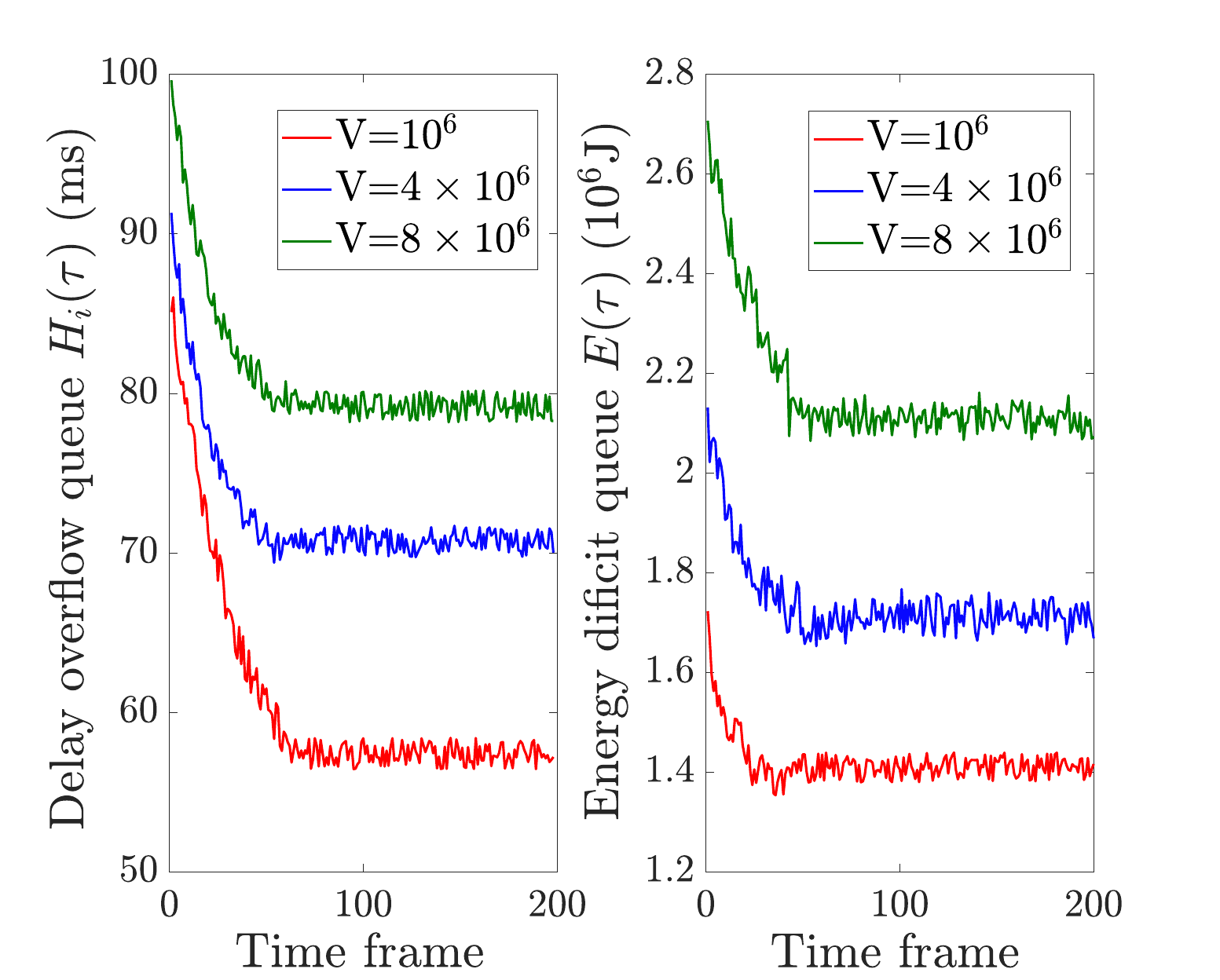}
}\hspace{-1.9em}
\subfigure[Queue backlogs w.r.t $K$.]{
\includegraphics[width=4.8 cm]{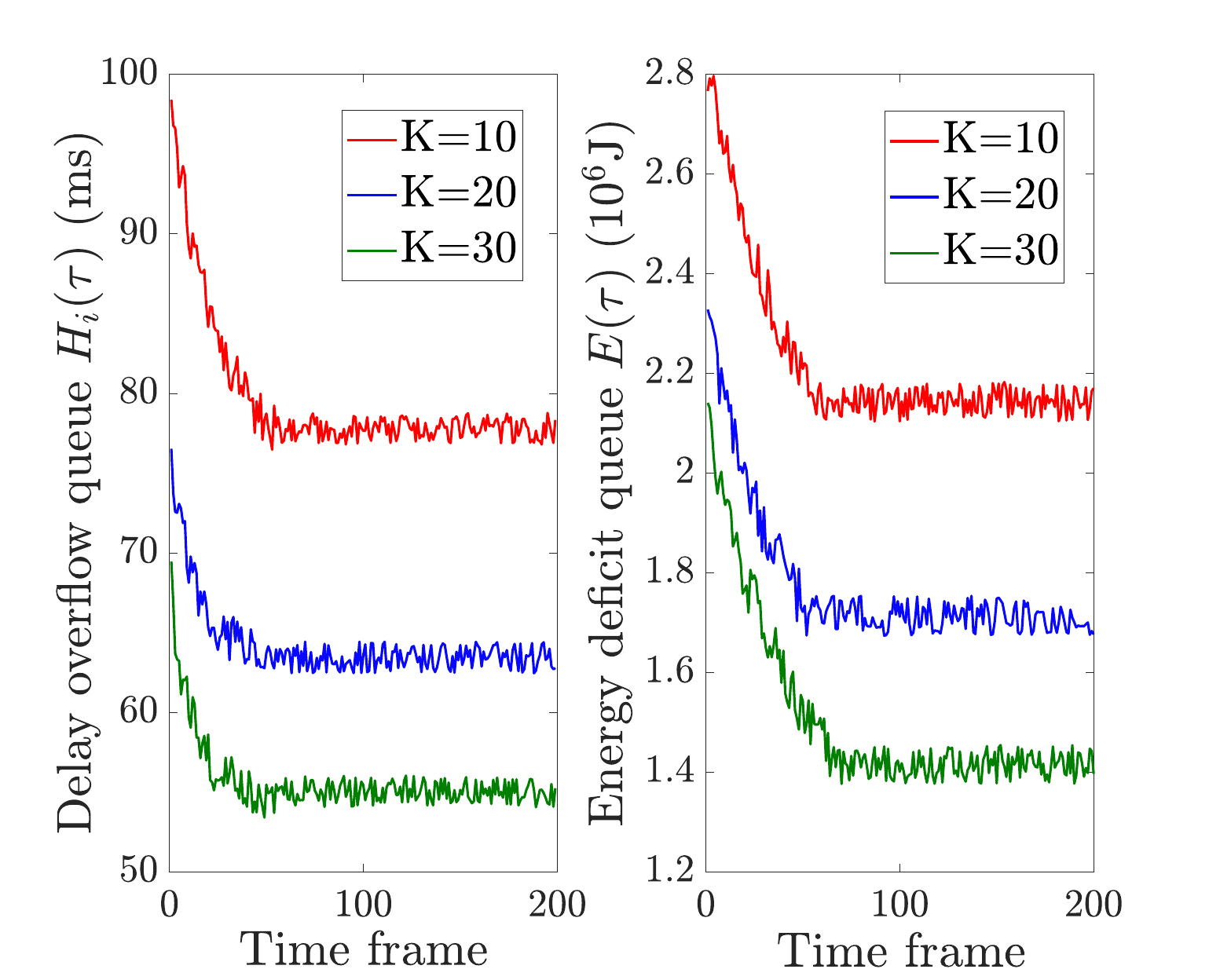}
}\hspace{-1.5em}
\caption{Stability of two queue backlogs by varying $V$ and $K$.}
\label{Queue}
\end{center}
\end{figure}

\begin{figure}[!t]
\begin{center}
\includegraphics[width=6.5 cm]{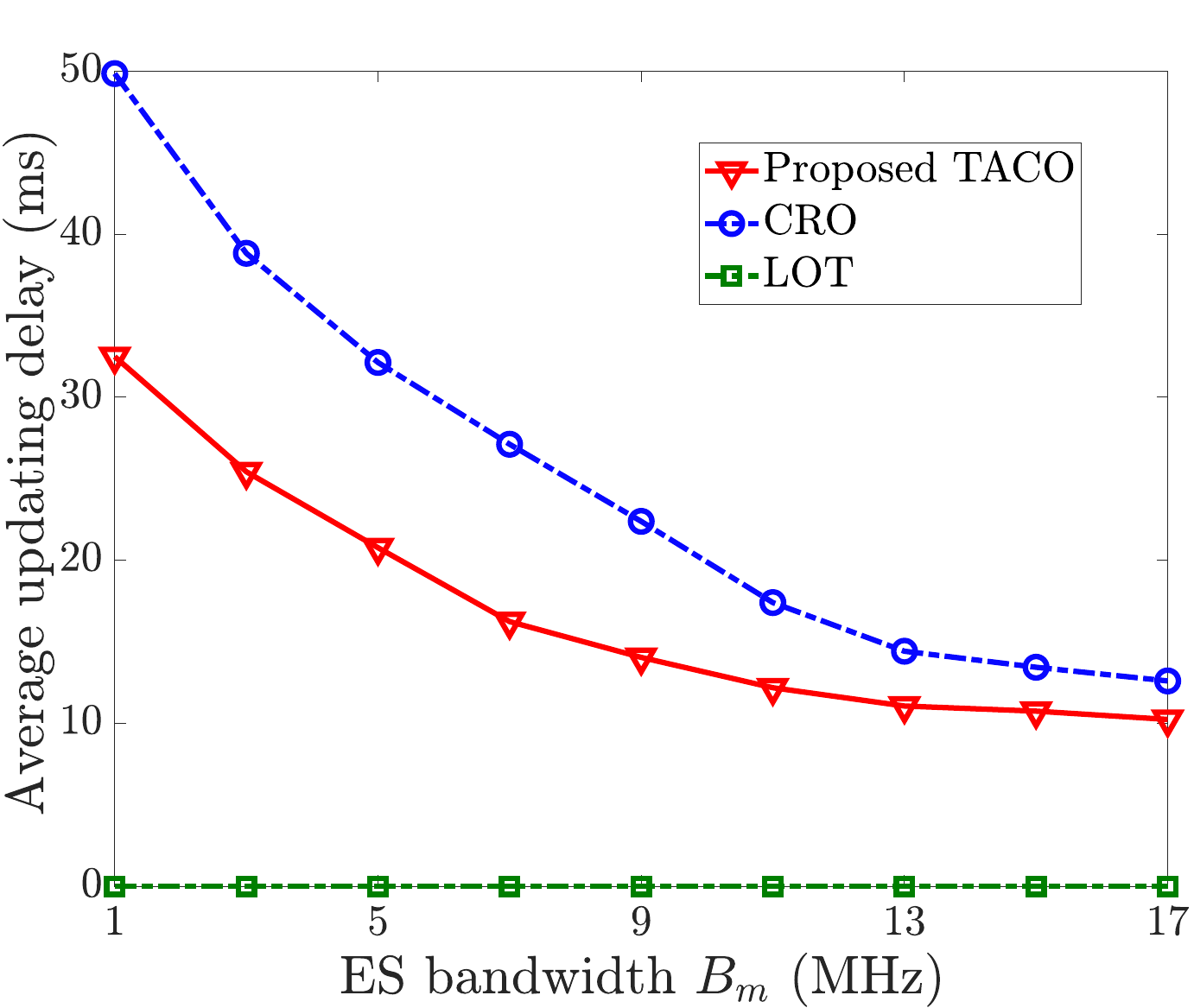}
  \caption{Comparison on updating delay of customized VT models with different bandwidth resources of ESs $B_m$.}
  \label{UpdatingDelay}
\end{center}
\end{figure}

\begin{figure}[!t]
\begin{center}
\includegraphics[width=6.5 cm]{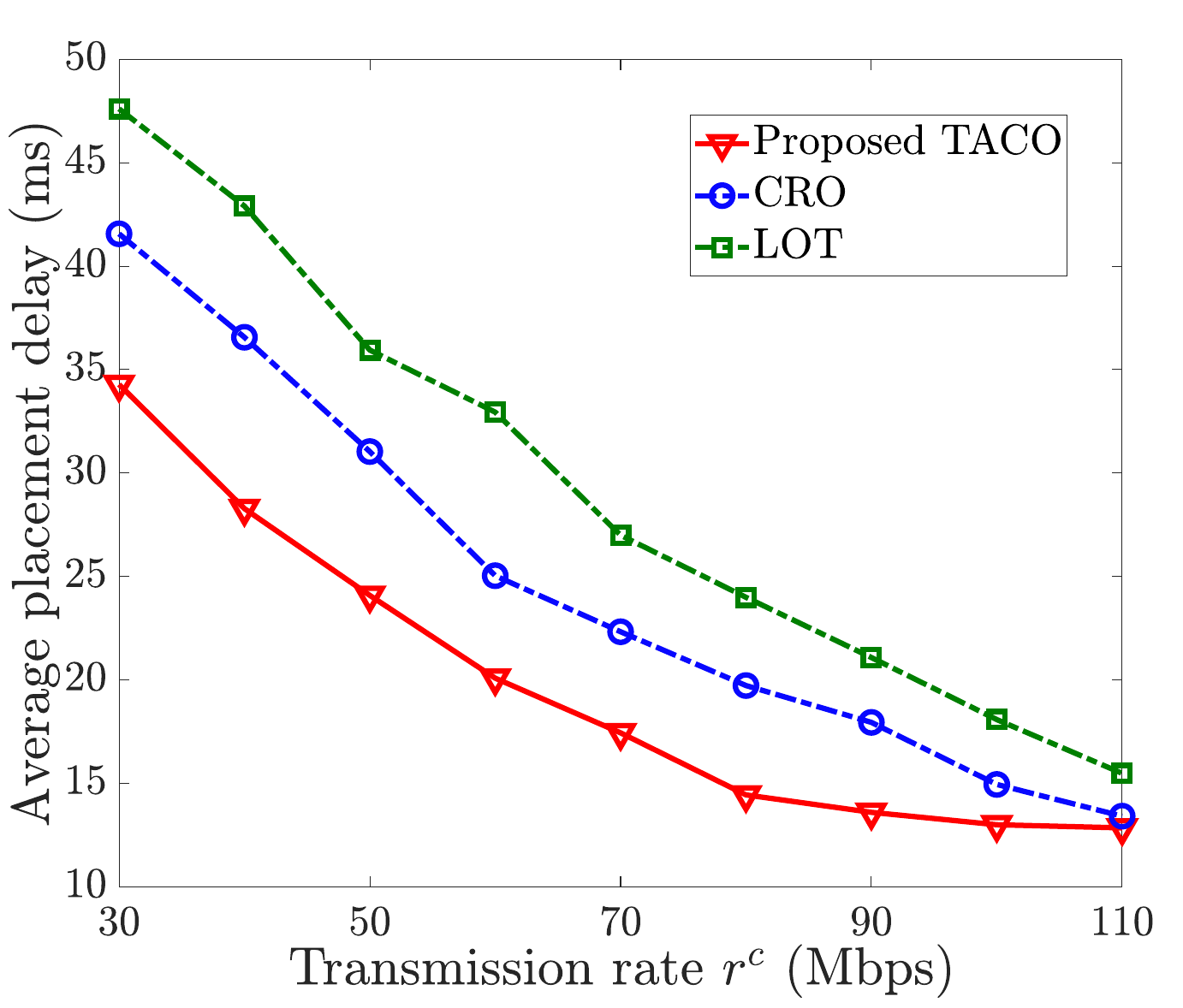}
  \caption{Comparison on placement delay of generic VT models with different transmission rate $r^c$.}
  \label{PlacementDelay}
\end{center}
\end{figure}

Fig. \ref{Queue} demonstrates the stability of the proposed TACO approach by showing the performance of delay overflow queue backlog $H_i(\tau)$ and energy consumption deficit queue backlog $E(\tau)$ brought by the Lyapunov decomposition by varying $V$ and $K$.
From this figure, we can see that two queue backlogs decrease and quickly stabilize over the time, because TACO focuses on controlling system costs (i.e., service response delay and system energy consumption) to minimize the objective function of $\mathcal{P}_3$, thereby shrinking two queue backlogs, and can eventually achieve a well balance between the task execution accuracy and system costs, leading to a stable outcome.
Besides, in Fig. \ref{Queue}(a), it is intuitive that both queue backlogs stabilize on higher values with the increase of $V$ as more emphasis is on maximizing the task execution accuracy, resulting in the growth of delay and energy consumption.
Meanwhile, Fig. \ref{Queue}(b) shows that both queue backlogs stabilize on lower values with the raise of $K$. This is because a larger value of $K$ means that generic VT models are placed with a lower frequency, and hence greatly reduces the generic VT model placement delay (i.e., $T_{i,m}^{dl}(\tau)+T_{i,m}^{pl}(\tau)$) and energy consumption (i.e., $E_{i,m}^{dl}(\tau)+E_{i,m}^{pl}(\tau)$).

\begin{figure}[!t]
\begin{center}
\includegraphics[width=6.5 cm]{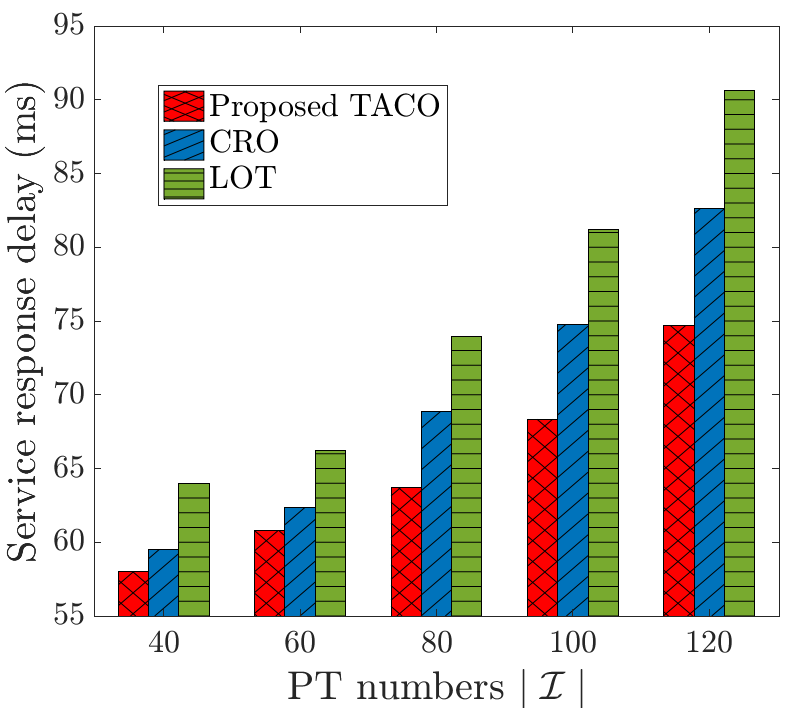}
  \caption{Comparison on the service response delay with different numbers of PT $I$.}
  \label{SRDelay}
\end{center}
\end{figure}

\begin{figure}[!t]
\begin{center}
\includegraphics[width=6.5 cm]{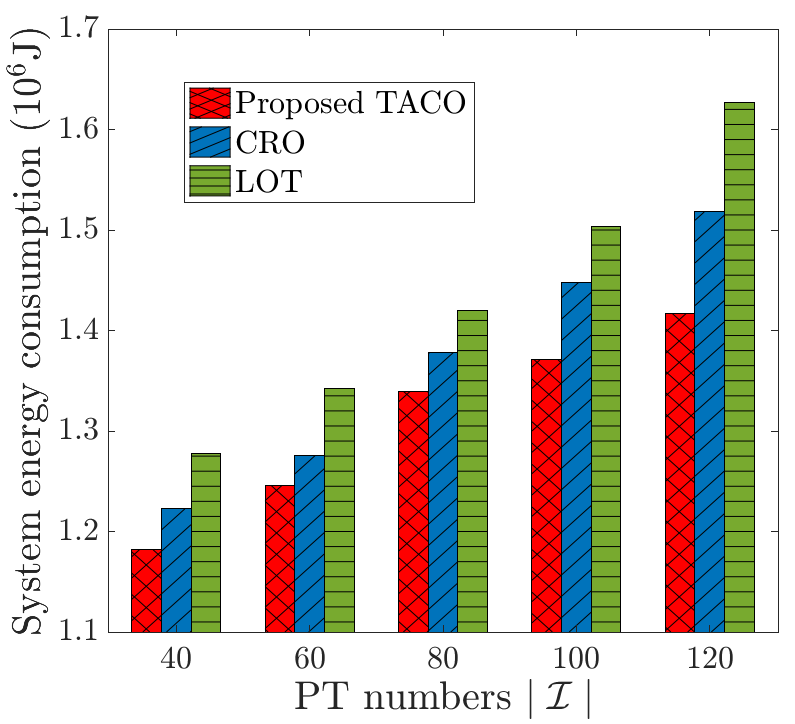}
  \caption{Comparison on the system energy consumption with different numbers of PT $I$.}
  \label{SystemEnergy}
\end{center}
\end{figure}

\begin{figure*}[!t]
\begin{center}
\subfigure[Accuracy w.r.t $B_m$.]{
\includegraphics[width=5.5 cm]{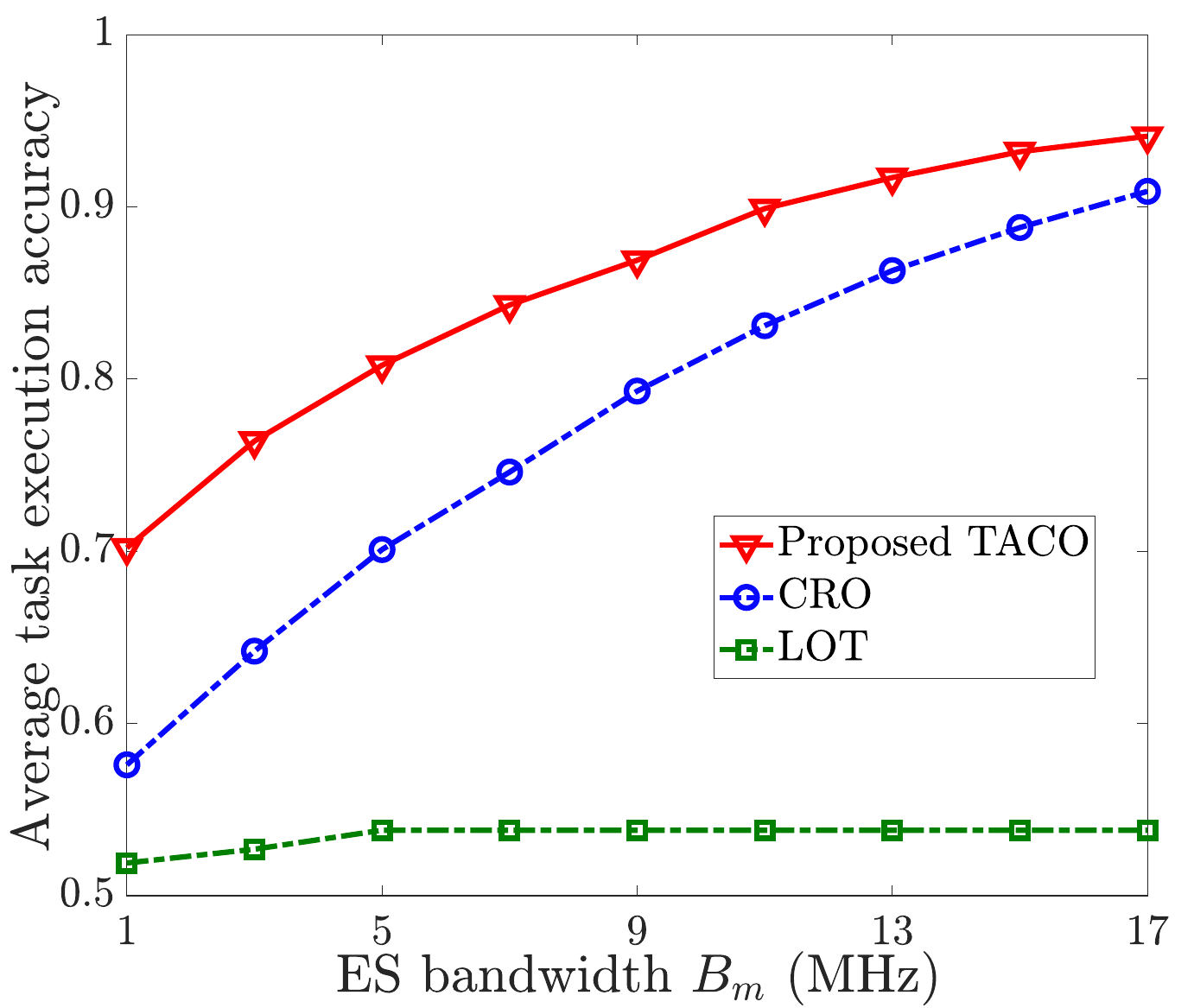}
}
\subfigure[Accuracy w.r.t $r^c$.]{
\includegraphics[width=5.8 cm]{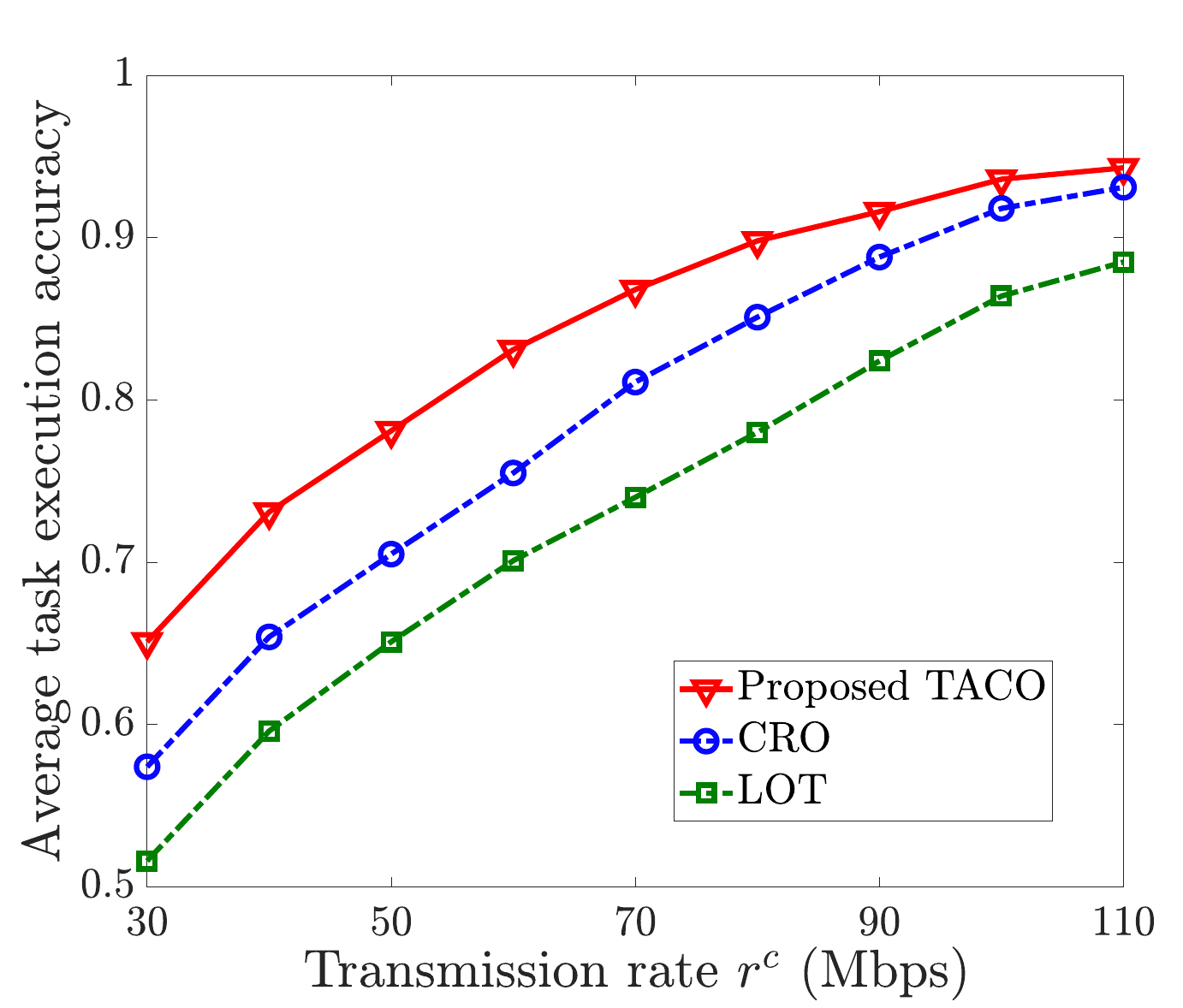}
}
\subfigure[Accuracy w.r.t $I$.]{
\includegraphics[width=5.8 cm]{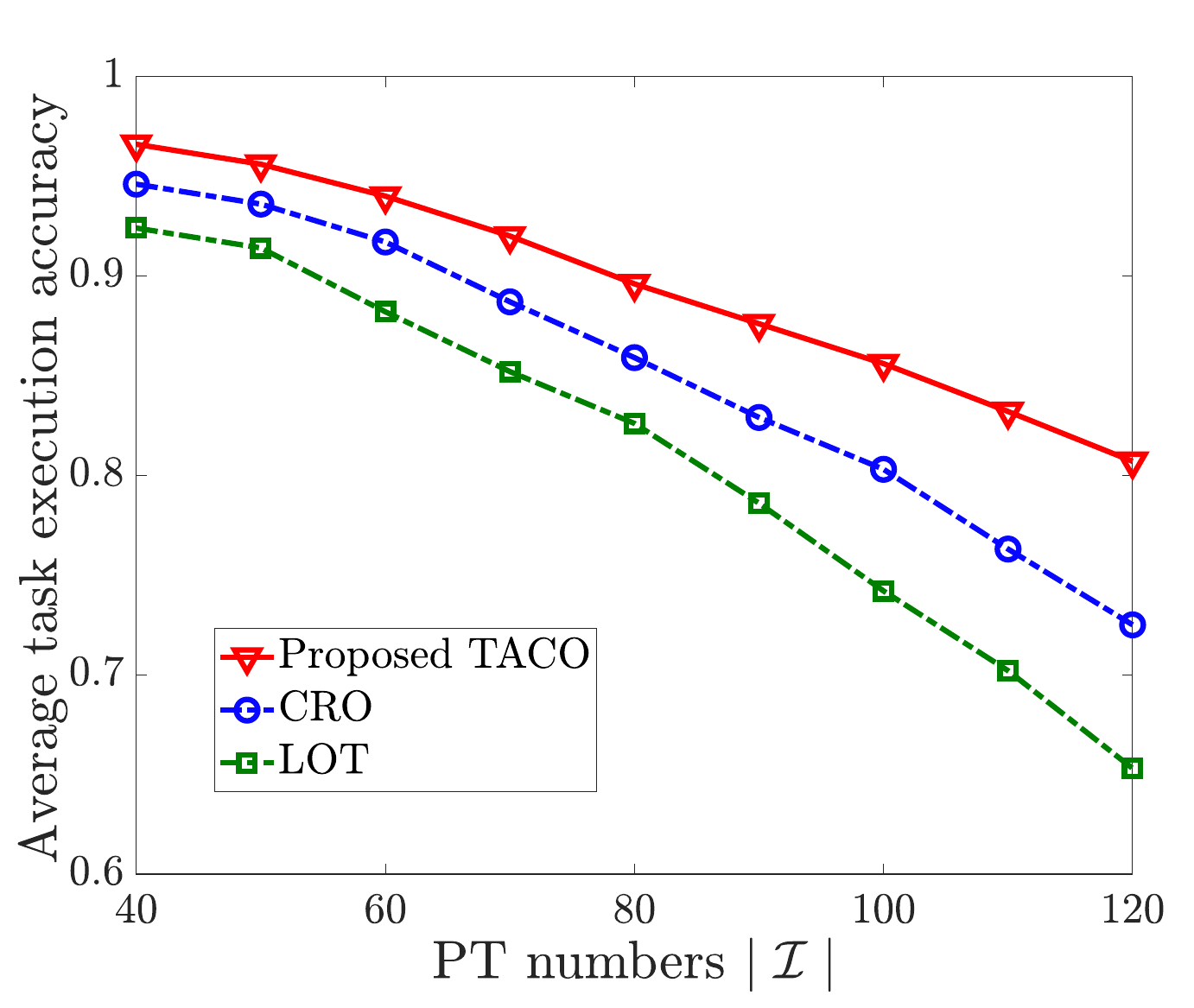}
}
\caption{Comparison on the average task execution accuracy by varying bandwidth resources of ESs $B_m$, transmission rate $r^c$ and numbers of PTs $I$.}
\label{AccuracyCompare}
\vspace{-1em}
\end{center}
\end{figure*}

Fig. \ref{UpdatingDelay} illustrates the average updating delay of customized VT models (i.e., $\sum_{t\in\mathcal{T}}\sum_{\tau\in\mathcal{T}_t}\sum_{i\in\mathcal{I}}(T_{i,m}^{ul}(\tau)+T_{i,m}^{ud}(\tau))/TKI$) with different ES bandwidth $B_m$ under LOT, CRO and the proposed TACO approach.
Since LOT ignores customized VT model update, meaning that personalized data of PTs does not need to be uploaded and processed, the average updating delay maintains as zero regardless of the ES's bandwidth $B_m$. In contrast, the average updating delay decreases with the increase of $B_m$ for both CRO and the proposed TACO approach due to the reduction in uploading delay $T_{i,m}^{ul}(\tau)$, as more uplink bandwidth resource is provided.
Besides, we can also see that TACO outperforms CRO because, with the help of generic VT model placement, TACO needs to upload much less personalized data than that of CRO.

Fig. \ref{PlacementDelay} shows the average placement delay of generic VT models (i.e., $\sum_{t\in\mathcal{T}}\sum_{\tau\in\mathcal{T}_t}\sum_{i\in\mathcal{I}}(T_{i,m}^{dl}(t)+T_{i,m}^{pl}(t))/TKI$) with different cloud-ES transmission rate $r^c$ under LOT, CRO and the proposed TACO approach.
It is obvious that the placement delay decreases logarithmically with the raise of $r^c$ for all schemes.
Furthermore, from this figure, we can see that LOT has the worst performance because it ignores customized model update and can only download as much experiential knowledge as possible to improve the average task execution accuracy, resulting in the highest average placement delay.
In contrast, both CRO and TACO outperform LOT due to the introduction of customized VT update, alleviating the burden on generic VT model placement.
Moreover, the proposed TACO approach achieves the best performance (with the placement delay reduced by $14.3\%$ and $25.5\%$ on average compared to CRO and LOT, respectively). This is because TACO builds VTs in two timescales (with large-timescale generic model placement and small-timescale customized model update) which only requires to download experiential knowledge at the beginning of each time frame significantly reducing the total data size in the process of model placement.


Figs. \ref{SRDelay} and \ref{SystemEnergy} investigate the service response delay and system energy consumption, achieved by LOT, CRO and the proposed TACO approach with different numbers of PTs $I$.
Both figures show that the service response delay and system energy consumption increase exponentially with $I$ for all schemes, because a larger $I$ implies more demands for VT construction with potentially larger amount of data in competing limited communication and computation resources.
Besides, LOT has the worst performances in these two metrics because its task execution accuracy only relies on the experiential knowledge downloaded in VT construction, and thus it has to download much larger amounts of experiential knowledge for guaranteeing a desired task execution accuracy.
Meanwhile, CRO and TACO outperform LOT, especially when $I$ becomes larger, which reveals the necessity of customized VT model update in addition to generic VT model placement.
Moreover, the proposed TACO approach achieves the best performance, and the reason follows the same as that in explaining Fig. \ref{PlacementDelay}.

Fig. \ref{AccuracyCompare} compares the average task execution accuracy achieved by LOT, CRO and the proposed TACO approach by varying ES's bandwidth $B_m$, transmission rate $r^c$ and number of PTs $I$.
It is shown that, i) in Fig. \ref{AccuracyCompare}(a), the average task execution accuracy increases with $B_m$ for all schemes, except for LOT; ii) in Fig. \ref{AccuracyCompare}(b), the average task execution accuracy increases with $r^c$ for all three schemes; and iii) in Fig. \ref{AccuracyCompare}(c), the average task execution accuracy decreases with $I$ for all three schemes.
The main reason is that, when more resource is supplied (i.e., $B_m$ and $r^c$ are large) or total resource demand is less competitive (i.e., $I$ is small), more experiential knowledge and personalized data can be transmitted for dynamic VT construction and more tasks can be offloaded for HDT-assisted edge execution, all contributing to the enhancement of task execution accuracy.
Furthermore, from this figure, we can see that LOT has the worst performance because it only considers generic VT model placement, and CRO is better than LOT thanks to the joint consideration of both generic VT model placement and customized VT model update.
Moreover, TACO achieves the best performance because it further strikes the balance of generic VT model placement and customized VT model update by conducting these two processes asynchronously in two different timescales.

\section{Conclusion}\label{Conclusion}

In this paper, the optimization of HDT deployment at the network edge has been studied. Particularly, aiming to maximize the accuracy of complex task execution assisted by HDT under various system uncertainties (e.g., random mobility and status variations), a two-timescale online optimization problem is formulated for jointly determining VTs' construction (including dynamic generic model placement and customized model update) and PTs' task offloading together with the management of access selection and corresponding communication and computation resource allocations. A novel solution approach, called TACO, is proposed, which decomposes the online problem into a series of deterministic ones and then leverages PME-based and BCD-based algorithms for solving different subproblems in different timescales alternately. Theoretical analyses and simulations show that TACO is superior compared to counterparts in the optimization of HDT deployment at the edge for not only improving the HDT-assisted task execution accuracy, but also reducing the service response delay and overall system energy consumption.

\bibliographystyle{IEEEtran}

\bibliography{PaperRef.bib}

\newpage

\appendices
\section{}\label{AppendixA}

By squaring both sides of the delay overflow queue in (\ref{DelayQueue}), we have
\begin{align}
& H_i^2(\tau+1)=\{[H_i(\tau)+T_i^{tol }(\tau)-T_i^{max }]^{+}\}^2 \\
& \leq H_i^2(\tau)+(T_i^{tol }(\tau)-T_i^{max})^2+2 H_i(\tau)(T_i^{tol }(\tau)-T_i^{max}).\nonumber
\end{align}

By subtracting $H_i^2(\tau)$ from both sides, dividing by 2 and summing up all inequalities for $i\in \mathcal{I}$, we have
\begin{align}\label{DelayBacklog}
& \frac{1}{2} \sum_{i\in\mathcal{I}}(H_i^2(\tau+1)-H_i^2(\tau)) \\
& \leq \frac{1}{2} \sum_{i\in\mathcal{I}}(T_i^{tol}(\tau)-T_i^{max})^2+\sum_{i\in\mathcal{I}} H_i(\tau) (T_i^{tol}(\tau)-T_i^{max}).\nonumber
\end{align}

Similarly, by squaring both sides of the energy deficit queue in (\ref{EnergyQueue}), we have
\begin{align}
& E^2(\tau+1)=\{[E(\tau)+E^{tol}(\tau)-E^{max}]^{+}\}^2 \\
& \leq E^2(\tau)+(E^{tol}(\tau)-E^{max})^2+2 E(\tau) (E^{tol}(\tau)-E^{max}).\nonumber
\end{align}

By subtracting $E^2(\tau)$ from both sides and dividing by 2, we have
\begin{equation} \label{EnergyBacklog}
\begin{aligned}
& \frac{1}{2}(E^2(\tau+1)-E^2(\tau)) \\
& \leq \frac{1}{2}(E^{tol}(\tau)-E^{max})^2+E(\tau) (E^{tol}(\tau)-E^{max}).
\end{aligned}
\end{equation}

Combining (\ref{DelayBacklog}) and (\ref{EnergyBacklog}), we have
\begin{equation}
\begin{aligned}\label{LyapunovDrift}
& \hspace{-4pt}L(\boldsymbol \Theta(\tau+1))-L(\boldsymbol\Theta(\tau)) \\
& \hspace{-4pt}=\hspace{-2pt}\frac{1}{2}\hspace{-2pt} \sum_{i\in\mathcal{I}}(H_i^2(\tau+1)\hspace{-2pt}-\hspace{-2pt}H_i^2(\tau))\hspace{-2pt}+\hspace{-2pt}\frac{1}{2}(E^2(\tau+1)\hspace{-2pt}-\hspace{-2pt}E^2(\tau)) \\
& \hspace{-4pt}\leq \hspace{-2pt}\frac{1}{2}\hspace{-2pt} \sum_{i\in\mathcal{I}}(T_i^{tol}(max)\hspace{-2pt}-\hspace{-2pt}T_i^{max})^2\hspace{-2pt}+\hspace{-2pt}\frac{1}{2}(E^{tol}(max )\hspace{-2pt}-\hspace{-2pt}E^{max})^2 \\
& \hspace{-4pt}+\hspace{-2pt}\sum_{i\in\mathcal{I}}\hspace{-2pt} H_i(\tau) (T_i^{tol}(\tau)\hspace{-2pt}-\hspace{-2pt}T_i^{max})\hspace{-2pt}+\hspace{-2pt}E(\tau) (E^{tol}(\tau)\hspace{-2pt}-\hspace{-2pt}E^{max}).
\end{aligned}
\end{equation}

Lastly, by adding $V\cdot \sum_{i\in\mathcal{I}}A_i(\tau)$ to both sides of (\ref{LyapunovDrift}) and taking the expectation of both sides of $\boldsymbol{\Theta}(\tau)$, inequality (\ref{Theorem1Equ}) can be derived.

\section{}\label{AppendixB}
For subproblem $\mathcal{P}_{5-1}$, the Lagrangian function is given by
 \begin{equation}
\begin{aligned}\label{Ptau1}
& \mathcal{L}_1(\boldsymbol{y})=\sum_{i \in \mathcal{I}} H_i(\tau)z_i(\tau) \sum_{m \in\mathcal M}(T_{i, m}^{ul}(\tau)+T_{i, m}^{ud}(\tau))\\
& +E(\tau)\sum_{i \in \mathcal{I}} \sum_{m \in \mathcal M} z_{i}(\tau)(E_{i, m}^{ul}(\tau)+E_{i, m}^{ud}(\tau))\\
& -V \cdot \sum_{i \in \mathcal{I}}A_i(\tau).
\end{aligned}
\end{equation}
By taking the first-order and second-order derivatives of (\ref{Ptau1}), we have
\begin{equation}
\begin{aligned}
& \frac{\partial \mathcal{L}_1}{\partial y_i(\tau)}= \sum_{i \in \mathcal{I}} H_i(\tau)z_i(\tau) \sum_{m \in \mathcal M} (\frac{a_{i, m}(t)S_i(\tau) C_m}{f_i(\tau) F_m}\\
& +\frac{S_i(\tau)}{r_{i, m}(\tau)})+E(\tau)\sum_{i \in \mathcal{I}} \sum_{m \in M} (a_{i, m}(t)\rho_m(F_m)^2 S_i(\tau) \\
& C_m+\frac{S_i(\tau) p_i}{r_{i, m}(\tau)})-2 V \sum_{i \in \mathcal{I}} \frac{z_i(\tau)S_i(\tau)}{D_i(t)+S_i(\tau)},
\end{aligned}
\end{equation}

\begin{equation}
\begin{aligned}
& \frac{\partial^2 \mathcal{L}_1}{\partial (y_i(\tau))^2}=2 V \cdot \sum_{i \in \mathcal{I}} S_i(\tau)^2 \geq 0.
\end{aligned}
\end{equation}
Since $\frac{\partial^2 \mathcal{L}_1}{\partial (y_i(\tau))^2} \geq 0$, we can conclude that problem $\mathcal{P}_{5-1}$ is convex.


For subproblem $\mathcal{P}_{5-2}$, we denote $\boldsymbol{\gamma}=\{\gamma_i(\tau),\forall i\in\mathcal{I},\forall \tau \in \mathcal{T}_t\}$ as the corresponding Lagrangian multiplier, which can be expressed as
\begin{equation}\label{57}
\begin{aligned}
& \mathcal{L}_2(\boldsymbol{b},\boldsymbol{\gamma})= \sum_{i \in \mathcal{I}} H_i(\tau) z_i(\tau) \sum_{m \in \mathcal{M}} (T_{i, m}^{ul}(\tau)+T_{i, m}^{ofld}(\tau))\\
& +E(\tau) \sum_{i \in \mathcal{I}} \sum_{m \in \mathcal{M}} (E_{i, m}^{ul}(\tau)+E_{i, m}^{ofld}(\tau)) \\
& + \sum_{i \in \mathcal{I}}\gamma_i(\tau)(a_{i,m}(t)b_{i}(\tau)-1).
\end{aligned}
\end{equation}
Since $1/r_{i,m}(\tau)$ in (\ref{57}) decreases monotonically, it is convex when $b_i(\tau) \in (0,1]$, and thus problem $\mathcal{P}_{5-2}$ is convex.


For subproblem $\mathcal{P}_{5-3}$, we denote $\boldsymbol{\eta}=\{\eta_i(\tau),\forall i\in\mathcal{I},\forall \tau \in \mathcal{T}_t\}$ as the corresponding Lagrangian multiplier, which can be calculated as
\begin{equation}
\begin{aligned}
& \mathcal{L}_3(\boldsymbol{f},\boldsymbol{\eta})= \sum_{i \in \mathcal{I}}  H_i(\tau)(\sum_{m\in\mathcal M}T_{i,m}^{pl}(t)/K +z_i(\tau)\sum_{m\in \mathcal M}  \\
& T_{i,m}^{ud}(\tau)+T_{i}^{exec}(\tau))+E(\tau)\sum_{i \in \mathcal{I}} (\sum_{m \in\mathcal M}E_{i, m}^{pl}(t)/K \\
& +\sum_{m \in\mathcal M}z_i(\tau)E_{i, m}^{ud}(\tau)+E_{i}^{exec}(\tau))\\
& + \sum_{i \in \mathcal{I}}\eta_i(\tau)(a_{i,m}(t)f_{i}(\tau)-1)
\end{aligned}
\end{equation}
Taking the first-order and second-order derivatives yields
\begin{equation}
\begin{aligned}
& \frac{\partial \mathcal{L}_3}{\partial f_i(\tau)}= -\sum_{i \in \mathcal{I}}  H_i(\tau)(\sum_{m\in\mathcal M}\frac{a_{i,m}(t)x_{i}(t)D_i(t) C_m}{(f_{i}(tK))^2F_m K}\\
& +z_i(\tau)\sum_{m\in \mathcal M}\frac{a_{i,m}(t)y_{i}(\tau)S_i(\tau) C_m}{(f_{i}(\tau))^2F_m}\\
& +\sum_{m\in\mathcal{M}}a_{i,m}(t) z_i(\tau)\frac{\lambda_i(\tau) C_m}{(f_{i}(\tau))^2F_m})\\
& + \sum_{i \in \mathcal{I}}\eta_i(\tau)a_{i,m}(t),
\end{aligned}
\end{equation}
\begin{equation}
\begin{aligned}
& \frac{\partial^2 \mathcal{L}_3}{\partial (f_i(\tau))^2}= 2\sum_{i \in \mathcal{I}}  H_i(\tau)\{\sum_{m\in\mathcal M}\frac{a_{i,m}(t)x_{i}(t)D_i(t) C_m}{(f_{i}(tK))^3F_m K}\\
& +z_i(\tau)\sum_{m\in \mathcal M}\frac{a_{i,m}(t)y_{i}(\tau)S_i(\tau) C_m}{(f_{i}(\tau))^3F_m}\\
& +\sum_{m\in\mathcal{M}}a_{i,m}(t) z_i(\tau)\frac{\lambda_i(\tau) C_m}{(f_{i}(\tau))^3F_m}\}.
\end{aligned}
\end{equation}
Obviously, when $f_i(\tau)\in(0,1]$, $\frac{\partial^2 \mathcal{L}_3}{\partial (f_i(\tau))^2} \geq 0$, and thus problem $\mathcal{P}_{5-3}$ is convex.


For subproblem $\mathcal{P}_{5-4}$, the Lagrangian function is given by
\begin{align}\label{61}
& \mathcal{L}_4(\boldsymbol{z})= \sum_{i \in \mathcal{I}}  H_i(\tau)[z_i(\tau)\sum_{m\in \mathcal M}(T_{i,m}^{ul}(\tau)+T_{i,m}^{ud}(\tau)  \nonumber\\
& +T_{i,m}^{ofld}(\tau))+T_{i}^{exec}(\tau)]+E(\tau)\sum_{i \in \mathcal{I}} [\sum_{m \in\mathcal M}z_i(\tau)(E_{i,m}^{ul}(\tau) \nonumber\\
& +E_{i, m}^{ud}(\tau)+E_{i, m}^{ofld}(\tau))+E_{i}^{exec}(\tau)]-V \sum_{i \in \mathcal{I}}A_i(\tau).
\end{align}
By taking the first-order derivative and second-order derivatives of (\ref{61}), we have
\begin{equation}
\begin{aligned}
& \frac{\partial \mathcal{L}_4}{\partial z_i(\tau)}= \sum_{i \in \mathcal{I}}  H_i(\tau)[\sum_{m\in \mathcal M}(T_{i,m}^{ul}(\tau)+T_{i,m}^{ud}(\tau)  \\
& +T_{i,m}^{ofld}(\tau))+T_{i}^{exec}(\tau)]+E(\tau)\sum_{i \in \mathcal{I}} [\sum_{m \in\mathcal M}(E_{i,m}^{ul}(\tau) \\
& +E_{i, m}^{ud}(\tau)+E_{i, m}^{ofld}(\tau))+E_{i}^{exec}(\tau)]\\
& -\hspace{-2pt} V\hspace{-2pt}\sum_{i\in\mathcal{I}}[1\hspace{-2pt}-\hspace{-2pt}(1\hspace{-2pt}-\hspace{-2pt}\frac{x_i(t)D_i(t)+y_i(\tau)S_i(\tau)}{D_i(t)+S_i(\tau)})^2\hspace{-2pt}-\hspace{-2pt}g_i^{local}].
\end{aligned}
\end{equation}
\begin{equation}
\begin{aligned}
& \frac{\partial^2 \mathcal{L}_4}{\partial (z_i(\tau))^2}= 0.
\end{aligned}
\end{equation}
Since $\frac{\partial^2 \mathcal{L}_4}{\partial (z_i(\tau))^2} \geq 0$, problem $\mathcal{P}_{5-4}$ is convex.
\end{document}